\documentclass[
  aps,
  prx,
  reprint,
  superscriptaddress,
  nofootinbib,
  floatfix,
  longbibliography
]{revtex4-2}

\usepackage{amsmath}

  \usepackage[T1]{fontenc}
  \usepackage{stix2}

  \makeatletter
  \newcommand\stix@tightfractions{%
    \check@mathfonts
    \fontdimen9\textfont2=0.3938\fontdimen6\textfont2  %
    \fontdimen10\textfont2=0.4437\fontdimen6\textfont2 %
    \fontdimen12\textfont2=0.3448\fontdimen6\textfont2 %
  }
  \AtBeginDocument{\begingroup
    \normalsize\stix@tightfractions
    \small\stix@tightfractions
    \footnotesize\stix@tightfractions
    \scriptsize\stix@tightfractions
  \endgroup}
  \makeatother

\usepackage{dsfont}
\usepackage{bm}
\usepackage{graphicx}
\usepackage{xcolor}
\usepackage{braket}
\usepackage{microtype}

\usepackage{tikz}
\usepackage{hyperref}
\definecolor{oklink}{HTML}{009E73}
\hypersetup{
  colorlinks = true,
  linkcolor  = black,
  citecolor  = oklink,
  urlcolor   = oklink,
  breaklinks = true,
}
\usepackage[nosort]{cleveref}
\crefname{equation}{Eq.}{Eqs.}
\Crefname{equation}{Equation}{Equations}
\crefname{figure}{Fig.}{Figs.}
\Crefname{figure}{Figure}{Figures}
\crefname{section}{Sec.}{Secs.}
\Crefname{section}{Section}{Sections}
\crefname{appendix}{Appendix}{Appendices}
\Crefname{appendix}{Appendix}{Appendices}

\makeatletter
\AtBeginDocument{\let\H@refstepcounter\refstepcounter}
\makeatother
\definecolor{okblue}{HTML}{0072B2}
\definecolor{okverm}{HTML}{D55E00}

\DeclareMathOperator{\E}{\mathbb{E}}
\DeclareMathOperator{\tr}{\mathrm{tr}}
\DeclareMathOperator{\Imr}{\mathrm{Im}}

\DeclareMathOperator{\Det}{\mathrm{Det}}
\DeclareMathOperator{\erfc}{\mathrm{erfc}}
\DeclareMathOperator{\erfi}{\mathrm{erfi}}
\newcommand{\bC}{\bar C}
\newcommand{\ph}{\hat\varphi}
\newcommand{\Drel}{\mathcal D_{\mathrm{rel}}}

\newcommand{\nchi}{\varsigma}

\newcommand{\Dperp}{D_\perp}
\newcommand{\Zc}{\mathbf Z}
\newcommand{\Ber}{\mathrm{Ber}}

\begin{document}

\title{Non-vanishing Density of States in Disordered Weyl Semimetals}
\author{Justin H. Wilson}
\affiliation{Department of Physics and Astronomy, Louisiana State University, Baton Rouge, Louisiana 70803, USA}
\affiliation{Center for Computation and Technology, Louisiana State University, Baton Rouge, Louisiana 70803, USA}
\date{September 24, 2026}

\begin{abstract}
There is disagreement in the literature concerning whether the nodal point of a three-dimensional Weyl semimetal survives weak short-range disorder: instanton calculations including fluctuations claim that the zero-energy density of states $\rho(0)$ vanishes, while exact numerics find a systematically finite number.
We settle this disagreement with a full saddle-point calculation, including both instanton and fluctuations.
Within the supersymmetric formulation for disorder, the instanton obeys a nonlinear Weyl equation whose solution is an exact $j=1/2$ hedgehog whose full functional form we numerically compute. 
We then compute the fluctuations about this saddle point, carefully identifying all zero modes, and computing the reduced superdeterminant as a convergent Fredholm determinant.
No fermionic zero mode beyond a Kramers doublet exists, and at finite disorder $w$, the fluctuations become a finite one-loop prefactor, $\rho(0)=\mathcal A w^{-4} \exp(-s^*/w^2)(1+O(w^2))$, in normalized units with $s^* = 6.4163(2)$ and $\mathcal A = 27.47(8)$ for Gaussian-correlated disorder.
This expression matches exact numerics on a single Weyl cone over four orders of magnitude with no fitted parameters (the amplitude is $\times 0.76$ the one-loop value), establishing that the density of states is finite for any disorder strength.
\end{abstract}

\maketitle

\section{Introduction and main result}
\label{ref:intro}

It has been suggested that the semimetallic point of a three-dimensional Weyl semimetal survives small-to-moderate disorder \cite{BuchholdAltland2018, BuchholdAltland2018a} despite numerical evidence to the contrary \cite{PixleyDasSarma2016, PixleyDasSarma2016a, WilsonPixley2020a, PixleyWilson2021}. 
However, the closest that the literature has gotten to a full refutation of this claim comes from a detailed analysis of Levinson's theorem along with numerical support consistent with the analytic form predicted by a finite density of exactly solvable defects \cite{PiresLopes2021}.
Rare-region arguments and instanton calculations \cite{NandkishoreSondhi2014} predict a nonzero nodal point density of states $\rho(0) \sim e^{-s^*/w^2}$ for dimensionless disorder strength $w$ below a perturbatively predicted phase transition \cite{Fradkin1986a, Fradkin1986, GoswamiChakravarty2011, RoyDasSarma2014, SyzranovGurarie2015, SyzranovRadzihovsky2018}, rounding the semimetal-to-diffusive metal transition \cite{SbierskiBrouwer2014, SbierskiBrouwer2015, Sbierski-2017, AltlandBagrets2015} into an avoided quantum critical point.
Contrariwise, Refs.~\cite{BuchholdAltland2018, BuchholdAltland2018a} claim that fermionic fluctuations force the instanton contribution to vanish identically (or be superexponentially suppressed).

\begin{figure}
  \includegraphics[width=\columnwidth]{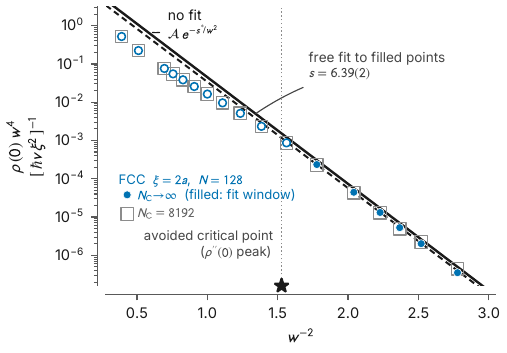}
  \caption{\textbf{Zero-energy density of states.}
    $\rho(0)\,w^4$ where $w = W\xi/\hbar v$ and $\rho$ is in units of $1/\hbar v\xi^2$. Data points from the single-cone calculations in Ref.~\cite{WilsonPixley2020a} ($k$-space fcc discretization, $\xi = 2a$, $N = 128$): blue circles are extrapolated in Chebyshev order, $N_{\mathrm C}\to\infty$; grey squares are the largest computed order, $N_{\mathrm C} = 8192$; error bars are smaller than the markers.
\Cref{eq:result} is the straight solid line with slope $-s^*$ with no fitted parameters.
The dashed line is a fit over the lowest six data points (filled): the exponent agrees to $1$--$2\%$ and the amplitude is $27\%$ below the analytic one-loop value ($\times0.73$).
Open circles are at or after the avoided quantum critical point (star and dotted vertical line found via $\rho''(0)$ peak, $w_c^{-2} \approx 1.53$). Not plotted: a one-parameter fit of the lowest six points letting $\mathcal A$ vary leading to an amplitude $24\%$ below the analytic one-loop value.
 }
  \label{fig:dos}
\end{figure}

In this work, we resolve this discrepancy and compute $\rho(0)$.
We work within the supersymmetric formulation of disorder which Refs.~\cite{BuchholdAltland2018,BuchholdAltland2018a} first used for this problem.
We develop a robust numerical technique to solve the saddle point of the disorder-averaged action for an instanton solution.
This instanton (the solution to a nonlinear Weyl equation) enables us to numerically compute $s^*$ as well as the fluctuation determinant.
With the exact enumeration of zero modes in hand --- six bosonic and four fermionic --- we employ supersymmetric field theory techniques to resolve their influence on the final result. 
Importantly, we demonstrate that no other zero (or near zero) fermionic modes exist to suppress the final result.
The main results of this work are analytic, while the use of numerics to find the instanton and compute the fluctuation determinant helps support our analytic claims and allows for near-exact agreement of $s^*$ with previous numerics using a different technique \cite{WilsonPixley2020a}, see \cref{fig:dos}.

The final result is
\begin{equation}
  \rho(0)
  = \frac{\mathcal A}{\hbar v\, \xi^2}
     \Big(\frac{\hbar v}{W\xi}\Big)^{4}
     \exp\!\Big[-s^*\Big(\frac{\hbar v}{W\xi}\Big)^{2}\Big]
     \big(1 + \mathcal O(w^2)\big),
  \label{eq:result}
\end{equation}
where $v$ is the Fermi velocity, $\xi$ is the disorder correlation length, $W$ is the disorder strength, and the constants $s^* = 6.4163(2)$ and $\mathcal A = 27.47(8)$ are computed for Gaussian-correlated potential disorder.
The dimensionless disorder strength is $w \equiv W\xi/\hbar v$.
This is the Nandkishore--Huse--Sondhi form \cite{NandkishoreSondhi2014} with instanton action \emph{and} the fluctuation prefactor.

\Cref{fig:dos} shows the comparison of this result with the exact numerics performed in Ref.~\cite{WilsonPixley2020a}.
The data is on an fcc momentum-space lattice (lattice spacing $2\pi/N a$) of Ref.~\cite{WilsonPixley2020a} at $\xi = 2a$, Chebyshev order extrapolated $N_{\mathrm C} \rightarrow \infty$ ($N_\mathrm{C} = 8192$, also shown).
It is plotted with the $w^{-4}$ prefactor divided out, so that \cref{eq:result} appears as a straight line of slope $-s^*$.
Nothing is fitted in the solid curve.
The measured exponent agrees with $s^*$ to $1$--$2\%$ (a free-slope fit over the lowest six points gives $s = 6.39\pm0.02$), and the amplitude sits $\approx 27\%$ below the one-loop value (factor $0.73$) --- the natural size of the $\mathcal O(w^2)$ correction at the accessible $w^2 \approx 0.4$--$0.6$ or of a systematic correction from the $k$-space discretization.
If we \emph{fix} the exponent and only fit $\mathcal A$, we arrive at a value $\approx 24\%$ below the one-loop value (factor $0.76$); we will use this one-parameter fit for discussions below.
The same window fitted \emph{without} the $w^{-4}$ prefactor returns $s = 5.4$--$5.6$, $13$--$16\%$ below $s^*$, indicating the power-law is necessary for agreement.

Field-theoretic methods, and in particular supersymmetry, for disordered problems have a long history.
In the replica formalism, Wegner formulated the mobility-edge problem (the Anderson transition) as one of continuous symmetry breaking; in this way, the retarded (or advanced) prescription acts as a small symmetry-breaking field leading to a finite lifetime in analogy to a small magnetic field picking a direction for a ferromagnet across that transition, even as the field goes to zero~\cite{Wegner1979}.
When retarded and advanced contributions are combined and averaged, Sch\"afer and Wegner showed that the symmetry is \emph{hyperbolic} symmetry consistent with a noncompact bosonic manifold (for our case with only one Green's function, the retarded contribution, convergence on this noncompact space implies we keep our endpoints fixed when we deform our path) \cite{SchaferWegner1980}.
With the bosonic scaffolding in place, Efetov constructed the supersymmetric functional integral for quenched disorder by noting that the fermionic determinant exactly cancels the bosonic determinant without the need for replicas \cite{Efetov1983}.
In parallel, Parisi and Sourlas uncovered a supersymmetry for random-field and stochastic problems along with a set of exact Ward identities that provide a dimensional reduction for functions of objects invariant under supersymmetric transformations \cite{ParisiSourlas1979, ParisiSourlas1982}; we will see that an identity like this applies exactly to the zero-mode sector of our instanton solution (\Cref{sec:zeromode}).

As with any contour integration, there are pitfalls when we deform our contours of integration when using steepest-descent methods in this formalism.
First, it is crucial that the (noncompact) integration domain for the bosonic sector keeps its endpoints intact (and avoids singularities and branch cuts --- though the quartic action keeps our exponent analytic, the fluctuation factor needs a more careful treatment along the deformation) \cite{SchaferWegner1980, Zirnbauer1986, Zirnbauer1996}.
We \emph{must} also be careful about boundary terms when introducing singular changes of variables \cite{Efetov1983, VerbaarschotZirnbauer1985, Zirnbauer1996}.
By the same token, the infinitesimal introduced by the retarded propagator requires special care near the real axis \cite{McKaneStone1981,Zirnbauer1986}.
Properly taken into account, these will help us obtain a numerically accurate result and simultaneously satisfy the Ward identities imposed by supersymmetry; the only other danger is a spontaneous breaking of the formal supersymmetry akin to a Parisi--Sourlas situation with true competing minima connected by unsaturated fermionic zeros \cite{ParisiSourlas1982, TissierTarjus2012}.
These subtleties need to be dealt with in this problem since resolving the issue of a vanishing density of states could well hinge on the exact measure, the integration contour, and the Ward identities of the zero modes.

We lay out the paper in the following way.
\Cref{sec:methods} gives the methodology for obtaining these results.
Then we get into the problem in \Cref{sec:model} where we set up the model and the supersymmetric method for disorder.
In \cref{sec:saddle}, we solve the saddle point for the instanton numerically.
Then, in \cref{sec:fluct}, we fully analyze the fluctuation action by identifying every bosonic and fermionic zero mode, and evaluating what remains as a finite fluctuation determinant.
\Cref{sec:zeromode} then treats the fermionic zero-mode sector and amplitude fluctuations by organizing the instanton in a supermatrix of collective coordinates; this allows for clear Ward identities showing that supersymmetry survives the instanton's inclusion, and completes the calculation, saturating the fermionic zero modes and leading to \cref{eq:result}.
\Cref{sec:conclusion} concludes and gives an outlook.
Details in the appendices include how to remove and account for the bosonic amplitude mode and the fermionic translation modes from the fluctuation determinant (\Cref{app:normtrans}), the ``no channel asymmetry'' theorem that removes a potential one-loop term from the final answer (\Cref{app:kappa}), and how the integration contour is handled along with phases in the final answer (\Cref{app:thimble}).

\section{Methodology}
\label{sec:methods}

In this section, we document how the results in this paper were found for transparency and ease of reproduction.
The analytic program came from trying to understand the details of the fluctuations in previous literature. 
It involved traditional analytic calculation, notes and ideas developed through conversation with generative AI tools, and adversarial checks by AI tools instructed to be hostile (adversarial) to the analytic conclusions. 
Symbolic analysis was assisted with the aid of computer algebra systems.
Numerical algorithms were also built with the help of generative AI tools and extensively verified for accuracy (see Data Availability statement at the end of the document).

In short, the formulation of the problem, the choice of what to compute, and the physical judgment were the author's while the mechanical volume was machine-assisted and independently verified.

\section{Model and supersymmetric formulation}
\label{sec:model}

The model is the three-dimensional Weyl Hamiltonian with potential disorder ($x$ is a spatial 3-vector),
\begin{equation}
  H = -i \hbar v\, \bm\sigma \cdot \nabla + V(x), \quad
  \E_V[V(x)V(y)] = W^2 \bC(x-y),
\end{equation}
with $\E_V[V] = 0$.
Much of the analysis does not depend on the particular form of $\bC(x)$; for numerical calculations we use the Gaussian form $\bC(x) = e^{-x^2/\xi^2}$.
Below, we will use $w$ throughout (effectively, $\hbar = v = \xi = 1$), and $W$ reappears only where full units are restored.
We collect some of the key notation in \Cref{tab:notation} (and in cases of constants with computed values with the Gaussian correlator, we have listed those values).

\begin{table}[t]
\caption{\textbf{Notation}. Some of the main variables (top block) and the constants which we compute for the Gaussian-correlator instanton (bottom block), with defining equations or sections.}
\label{tab:notation}
\begin{ruledtabular}
\footnotesize
\setlength{\tabcolsep}{2pt}
\begin{tabular}{lll}
variable & description & defined \\
\colrule
$\rho(\omega)$ & density of states & \cref{eq:dos} \\
$w$, $W$ & disorder strength & \cref{eq:result} \\
$\xi$, $\bC(x)$ & correlation length; correlator & \cref{sec:model} \\
$K = i\bm\sigma\cdot\nabla + U$ & saddle operator; well & \cref{eq:saddle} \\
$\ph_{\pm1/2}$, $\hat n$ & normalized instanton; density & \cref{eq:saddle} \\
$A = \sqrt\gamma/w$ & instanton amplitude & \cref{eq:amplitude} \\
$p$ & norm-mode coordinate & \cref{sec:supermatrix} \\
$q = p + \nu$, $d$ & supermatrix invariants & \cref{sec:supermatrix} \\
$\mathcal H = H_0 + d\,\varkappa$ & fiber density; form factor & \cref{eq:Hweight} \\
$\Lambda_\pm(p)$ & fermion-pair responses & \cref{eq:Lambdadef} \\
$\Drel$ & fluctuation determinant & \cref{eq:Drel} \\
\colrule
$\gamma = 12.116035$ & self-consistent coupling & \cref{eq:saddle} \\
$s^* = 6.4163(2)$ & reduced action, $S_I = s^*/w^2$ & \cref{eq:sstar} \\
$n_t = 0.6723132(24)$ & translation-mode norm & \cref{sec:bosons} \\
$\nchi^2 = 0.185$ & rotation--translation overlap & \cref{sec:bosons} \\
$\Lambda_\pm(1) \!= \! -2.2835(49)$ & channel response & \cref{sec:kappa} \\
$\varkappa_1 = 0$ (exact) & channel asymmetry & \cref{eq:matching} \\
$\ln\Drel = 0.130(3)$ & $\Drel \approx 1.14$ & \cref{eq:Drelnum} \\
$\mathcal A = 27.47(8)$ & one-loop prefactor & \cref{eq:assembly} \\
\end{tabular}
\end{ruledtabular}%
\end{table}

Using $\int_x\equiv \int d^3 x$, we will use the standard functional inner product to simplify expressions
\begin{equation}
  \label{eq:innerprod}
  \braket{f, g} \equiv \int_x f(x) g(x),
\end{equation}
where if $f$ and $g$ have spinor indices, they are understood to also be contracted.

We begin with the disorder-averaged retarded Green's function $G^+_{\omega,x,x'} = \E_V \braket{x | (\omega + i0^+ - H)^{-1} | x'}$ and its relation to the density of states,
\begin{equation}
  \rho(\omega) = -\frac{1}{\pi L^3} \int_x \Imr \tr G^+_{\omega,x,x}.
  \label{eq:dos}
\end{equation}
To help us perform the average, we use supersymmetry \cite{Efetov1983}.  
We therefore introduce the superfield $\psi_x = (\phi_x, \chi_x)^T$, with $\phi_x$ bosonic and $\chi_x$ fermionic, and the action
\begin{equation}
  S_V[\psi] = \int_x \bar\psi_x\, (\omega + i0^+ - H)\, \psi_x,
  \label{eq:action}
\end{equation}
acting diagonally on both components.
We will use the bar to denote the conjugate field, and this notation collides with complex conjugation for bosons $\bar\phi = \phi^*$, whereas $\bar\chi$ is an independent Grassmann generator and not the complex conjugate of $\chi$.
Therefore, $z^*$ will indicate complex conjugation acting on a number $z$ in $\mathbb C$ and for Grassmann numbers $\chi^* =\chi$, leaving them unchanged; the distinction will matter in \cref{app:kappa}.
The action can be used in a few ways, which will be crucial to our analysis
\begin{enumerate}
  \item \emph{Partition Function}: for every disorder realization $V$,
\begin{equation}
  Z = \int \mathcal D[\bar\psi,\psi]\, e^{i S_V[\psi]} = 1,
  \label{eq:anchor}
\end{equation}
the fermionic determinant canceling the bosonic determinant mode-by-mode which we will use both exactly (for Ward identities) and approximately (far from the instanton core where bosons and fermions will match again).
Maintaining this is \emph{crucial} to obtaining the correct result for $\rho(0)$.
\item \emph{The Green's function}: with $\mathsf k = \operatorname{diag}(1,-1)$ in superspace, the graded bilinear recovers
\begin{equation}
  G^+_{\omega,x,x'} = -\frac{i}{2}\, \E_V \int \mathcal D[\bar\psi,\psi]\,
    (\bar\psi_{x'} \mathsf k \psi_x)\, e^{i S_V[\psi]},
  \label{eq:insertion}
\end{equation}
the boson and fermion loops contributing equally under $\mathsf k$ --- whence the counting factor $\tfrac12$, absorbed in Refs.~\cite{BuchholdAltland2018, BuchholdAltland2018a} into a $\tfrac{1}{2\pi}$ in \cref{eq:dos}.
\end{enumerate}

A related useful fact is that the bilinear without $\mathsf k$ must vanish for all realizations $V(x)$
\begin{equation}
  \int \mathcal D[\bar\psi,\psi]\,
    (\bar\psi_{x'} \psi_x)\, e^{i S_V[\psi]} = 0.
  \label{eq:notau3}
\end{equation}

Our disordered potential enters as $-i\int_x V_x n_x$ in the exponent and is coupled to the superdensity $n_x = \bar\psi_x \psi_x$.
Averaging this quantity with Gaussian disorder can be done exactly, $\E_V e^{-i\int Vn} = \exp[-\tfrac{w^2}{2}\int_{x,y} \bC(x-y) n_x n_y]$.
Following Ref.~\cite{BuchholdAltland2018a}, it is convenient to rotate the superfield, $(\psi,\bar\psi) = e^{i\pi/4}(\psi',\bar\psi')$; once rotated, we will immediately relabel $\psi' \mapsto \psi$ (and the same for $\bar\psi$).  %
On the rotated contour the exponentiated action is $e^{-E[\psi]}$ with 
\begin{equation}
  E[\psi] = \int_x \bar\psi\, (i\bm\sigma\cdot\nabla)\, \psi
  - \frac{w^2}{2} \int_{x,y} \bC(x-y)\, n_x n_y.
  \label{eq:E}
\end{equation}
This rotated expression will allow for an easy instanton solution but at the cost of a constant (phase) Jacobian which we account for later.
Note that the quartic enters the exponent $-E$ with a \emph{plus} sign and that the rotation's Jacobian on the zero-mode sector is a phase we account for below. 
The contour-level justification (the Lefschetz thimble through the saddle) is given in \cref{app:thimble}.

\section{The instanton saddle point}
\label{sec:saddle}

\begin{figure}
  \includegraphics[width=\columnwidth]{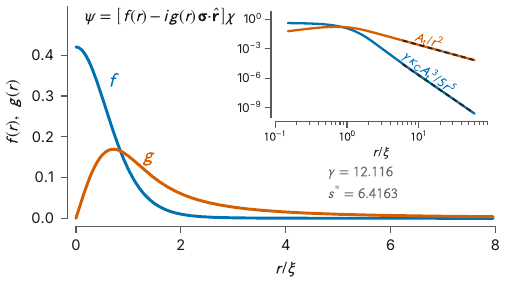}
  \caption{\textbf{The instanton.}
Radial profile of the $j = \tfrac12$ hedgehog solution of \cref{eq:saddle} for the Gaussian correlator, with self-consistent coupling $\gamma = 12.116$ and action $s^* = 6.4163$.
Inset: the same plot on logarithmic axes taken out to $r = 60\,\xi$ to show the asymptotic tails $g \to A_t/r^2$ and $f \to \gamma \kappa_C A_t^3/5r^5$ (dashed).
}
  \label{fig:inst}
\end{figure}

At $\omega = 0$ the saddle point lies in the bosonic sector, $\psi_I = (\varphi_I, 0)^T$.
Writing $\varphi_I = A\ph$ with $\int_x |\ph|^2 = 1$, steepest descent of \cref{eq:E} gives the nonlinear Weyl equation
\begin{equation}
  K \ph = 0, \quad
  K = i\bm\sigma\cdot\nabla + U, \quad
  U = -\gamma\, (\bC * \hat n),
  \label{eq:saddle}
\end{equation}
where $\hat n = |\ph|^2$ is the density of the normalized instanton, $\bC * \hat n \equiv \int_y \bC(x-y)\hat n(y)$ is the convolution of $\bC$ and $\hat n$ in real space, and the amplitude is exactly
\begin{equation}
  A^2 = \gamma / w^2.
  \label{eq:amplitude}
\end{equation}
The self-consistent value $\gamma$ is a generalized eigenvalue of \cref{eq:saddle} (and hence $w$-independent). 
We therefore tune $\gamma$ such that there is a normalizable zero-energy solution. 
As a result, all $w$ dependence resides in the amplitude $A$.
We solve \cref{eq:saddle} in the $j = \tfrac12$ channel with the hedgehog ansatz $\ph = [f(r) - i g(r)\, \bm\sigma\cdot\hat{\mathbf r}\,]\chi_\uparrow$, where $\chi_\uparrow$ is the constant spin-up two-spinor, by a self-consistent homotopy method described below, obtaining $\gamma = 12.116035$ [\cref{fig:inst}]; the profile carries the power-law tail $g \sim A_t/r^2$, $\hat n \sim A_t^2/r^4$ predicted in Ref.~\cite{NandkishoreSondhi2014}, confirming that analysis in the fully self-consistent nonlinear regime.
The solution is one member of a Kramers doublet: the hedgehog $\ph_{+1/2}$ has a partner $\ph_{-1/2}$ with the magnetic quantum number flipped,
\begin{equation}
  \begin{aligned}
  \ph_{+1/2} &= \phantom{-}\big[f(r) - i g(r)\, \bm\sigma\cdot\hat{\mathbf r}\,\big]\chi_\uparrow,\\
  \ph_{-1/2} &= -\big[f(r) - i g(r)\, \bm\sigma\cdot\hat{\mathbf r}\,\big]\chi_\downarrow,
  \end{aligned}
  \label{eq:doublet}
\end{equation}
and the doublet $\ph_m$, $m = \pm1/2$, satisfies $\ph_m^\dagger\ph_{m'} = \hat n\,\delta_{mm'}$ \emph{pointwise}.
Both members solve \cref{eq:saddle} at the same $\gamma$, and the pointwise identity is what organizes the zero-mode ledger of \cref{sec:bosons,sec:fermions} and the supermatrix of \cref{sec:supermatrix}.

While this solution describes one Kramers pair of instantons, there is another pair with the same numerical contribution.
Parity flips the Weyl term but not the potential, so $\ph(-x)$ solves \cref{eq:saddle} with $\gamma \to -\gamma$. 
This changes the instanton by exactly $(f, g) \to (f, -g)$, leaving $\hat n$, and therefore the instanton action \cref{eq:sstar}, unchanged.
Physically the pair is forced by the symmetry of the disorder distribution: a rare potential \emph{barrier} $-U$ is exactly as probable as the well $U$, and at a Weyl node it binds a zero-energy state just as well as the \emph{well}\footnote{Pun only slightly intended.}.
In the field theory, this parity-matched solution cannot be deformed with rotations, translations, or phase since those preserve the sign of $U$; in this case the amplitude $A^2=\gamma/w^2<0$ is related to the original contour by parity \cref{app:thimble}. 
This results in a simple multiplicity of $2$ for the final result that we restore in the assembled prefactor of \cref{sec:assembly}.

We can use \cref{eq:saddle} with the Weyl kinetic energy to obtain a virial-like identity, $\braket{\varphi_I, (i\bm\sigma\cdot\nabla)\varphi_I} = w^2 A^4 \braket{\hat n, \bC * \hat n}$ implying that the on-shell action can be written purely in terms of the energy of the self-consistent potential 
\begin{equation}
  S_I = E[\psi_I] = \frac{s^*}{w^2}, \qquad
  s^* = \tfrac12 \gamma^2 \braket{\hat n, \bC * \hat n},
  \label{eq:sstar}
\end{equation}
and the leading weight is $e^{-s^*/w^2}$.
For the Gaussian-correlated disorder we find $s^* = 6.4163(2)$ numerically (the uncertainty is limited by initial configurations, grid sizes, and cutoff distances, all of which we vary).
At this point, we already have a parameter-free fit from the slope in \cref{fig:dos} at the $1$--$2\%$ level with $s^*$ if we assume the $w^{-4}$ prefactor which we will derive in \cref{sec:assembly}.

\subsection{The self-consistent homotopy method}
\label{sec:homotopy}

For $j = \tfrac12$ and $\omega=0$, \cref{eq:saddle} reduces to a one-dimensional problem
\begin{equation}
  f' = U g, \qquad g' + \frac{2}{r} g = -U f,
  \label{eq:radial}
\end{equation}
with $U$ fixed by \cref{eq:saddle} and $\hat n = f^2 + g^2$, normalized such that $4\pi\int r^2 \hat n\, dr = 1$. 
The solution should be regular at the origin and have the algebraic tail $g \sim A_t/r^2$ to be well-defined and normalizable. 
As a result, $f \sim \gamma\kappa_C A_t^3/5r^5$, with $\kappa_C \equiv \int d^3x\, \bC(x) = \pi^{3/2}\xi^3$. 
We will use matching these asymptotic forms in place of a boundary condition at infinity.
The angular integral of the Gaussian correlator is
\begin{equation}
  C_0(r,r') = \frac{\pi\xi^2}{r r'}
  \Big[e^{-(r-r')^2/\xi^2} - e^{-(r+r')^2/\xi^2}\Big],
\end{equation}
which only depends on the two radii $r'$ and $r$.
The potential in \cref{eq:saddle} is $U = -\gamma\, Q[\hat n]$ with
\begin{equation}
  Q[\hat n](r) = \int_0^\infty dr'\, r'^2\, C_0(r,r')\, \hat n(r'),
  \label{eq:Qconv}
\end{equation}
which is a single dense matrix given a discretization of $r$.
All angular dependence has dropped out (due to $U = U(r)$) so that different $j$ sectors decouple.
And the nonlinearity lives in $Q$ which we solve self-consistently while solving the radial differential equations directly.

To solve \Cref{eq:radial}, we employ a nonstandard self-consistent method.
First, we should note that it is a nonlinear \emph{generalized eigenvalue} problem with eigenvalue $\gamma$.
A generic well depth does not admit a zero energy solution, so $\gamma$ is tuned (along with the shape of $\hat n$, the self-consistency). 
The obvious fixed-point iteration has many nearby minima; for instance, a seed of $\hat n = 0$ yields an effective plane-wave solution. 
And yet at the physical, computed coupling $\gamma \approx 12$ the nonlinearity iteration from a rough guess can be difficult to reliably converge.
We therefore deform from the square well that was initially solved in Ref.~\cite{NandkishoreSondhi2014} to guide the solution to the instanton.
We replace $U$ with $U_\delta$ defined through
\begin{equation}
  \Phi_\delta[\hat n] = (1-\delta)\,\Phi_{\mathrm{well}} + \delta\, Q[\hat n],
  \qquad
  U_\delta = -\gamma_\delta\, \Phi_\delta,
  \label{eq:homotopy}
\end{equation}
and the homotopy moves $\delta$ from $0$ to $1$, so that $U_{\delta=1} = U$.
The starting potential is the correlator-smeared spherical well $\Phi_{\mathrm{well}} = Q[n_{\mathrm{box}}]$, with $n_{\mathrm{box}}(r) = \tfrac{3}{4\pi R_w^3}\,\Theta(R_w - r)$ a square well normalized to integrate to one with radius $R_w = 1.5\,\xi$ standing in for the instanton density.
The end result does not retain any imposed potential at $\delta = 1$.

The solver is organized as three nested loops.
The outer loop takes $\delta$ through a set of increasing values, seeding each solve with the previous converged state; a step that fails to converge is bisected by inserting the midpoint value of $\delta$ and trying again.
The middle loop is the self-consistency iteration at fixed $\delta$, a damped Picard update $\hat n \leftarrow (1-\alpha)\,\hat n + \alpha\, \hat n_{\mathrm{new}}$ with mixing parameter $\alpha = 0.1$.

Each iteration of the self-consistent loop therefore yields a new well $\Phi_\delta$, and $\gamma_\delta$ must be recomputed so that this well supports a normalizable zero-energy solution.
For a trial $\gamma$ we integrate \cref{eq:radial} outward from the regular solutions at the origin, $f = 1 - U(0)^2 r^2/6$ and $g = -U(0)\, r/3$ ($r\ll \xi$), across a grid spanning $r \in [10^{-4}\xi, R]$ with $R = 60\,\xi$.
The self-consistent well after even one iteration decays algebraically $\Phi_\delta \sim \delta\,\kappa_C A_t^2/r^4$, but fast enough that the zero mode is normalizable (with $1/r^2$ power-law in the large $r$ asymptotics).
The naive box condition $f(R) = 0$ fails because of the power-law nature of the state; normalizability instead demands that $f$ join the analytic tail quoted below \cref{eq:radial}, which becomes the finite-$R$ matching condition
\begin{equation}
  M(\gamma) \equiv f(R) - \frac{\gamma\,\delta\,\kappa_C A_t^3}{5R^5} = 0,
  \qquad
  A_t = R^2 g(R),
  \label{eq:mismatch}
\end{equation}
with $A_t$ determined by the second equation (i.e., no fitting).
The root of $M$ is bracketed and bisected, keeping the lowest positive $\gamma$ whose profile is nodeless (the ground-state).
The matched boundary value is tiny ($f(R) \sim 10^{-10}$ at $R = 60\,\xi$), so for precision, the integration tolerances must be able to resolve these values (we use an adaptive fifth-order Runge--Kutta scheme at absolute tolerance $10^{-11}$).
And the tail carries real probability: $g \sim A_t/r^2$ places the weight $4\pi A_t^2/R$, about $1.3\%$ at $R = 60\,\xi$, beyond the grid, which the normalization $4\pi\int r^2\hat n\, dr = 1$ must include.

Convergence is certified by the weighted residual $\|U + \gamma Q[\hat n]\|/\|U\| < 10^{-8}$ (a diagnostic independent of the Picard mixing), the measured $r^{-2}$ tail slope, and branch independence: initial wells of radius $R_w = 0.75\,\xi$ and $1.5\,\xi$ give $s^*$ in agreement to $1.5\times10^{-7}$.
The instanton action is then computed from the converged profile, $s^* = \tfrac12\gamma^2\big(I_{\mathrm{core}} + I_{\mathrm{tail}}\big)$.
The overlap in \cref{eq:sstar} is one-dimensional, so splitting it at the grid edge $R$,
\begin{equation}
  \braket{\hat n, \bC * \hat n}
  = 4\pi\!\int_0^\infty\! dr\, r^2\, \hat n(r)\, Q[\hat n](r)
  \equiv I_{\mathrm{core}} + I_{\mathrm{tail}},
  \label{eq:Isplit}
\end{equation}
where $I_{\mathrm{core}} = 4\pi\int_0^R dr\, r^2\, \hat n\, Q[\hat n]$ (numerically evaluated with the same grid and quadrature as the normalization), and $I_{\mathrm{tail}} = 4\pi \int_R^\infty dr \, r^2 \, \hat n \, Q[\hat n]$.
Asymptotically, $\hat n \simeq A_t^2/r^4$ for $r\gg \xi$, and therefore $Q$ acts as $\kappa_C\,\delta^{(3)}$, a Dirac delta. 
In other words, $\bC * \hat n \to \kappa_C \hat n$ with the closed form,
\begin{equation}
  I_{\mathrm{tail}} = 4\pi\kappa_C A_t^4\!\int_R^\infty\! \frac{dr}{r^6} = \frac{4\pi\kappa_C A_t^4}{5R^5},
  \label{eq:Itail}
\end{equation}
with $A_t$ determined by the profile \cref{eq:mismatch}.
At $R = 60\,\xi$, $I_\mathrm{tail}$ provides a correction below the quoted uncertainty on $s^*$, but is nonetheless a check that we are not missing a contribution in the asymptotic tail.

\section{Fluctuations}
\label{sec:fluct}

From this point forward, the computation of saddle-point fluctuations takes the shape of a traditional semiclassical expansion.
The classical action is controlled by the instanton amplitude $A = \sqrt\gamma/w$ of \cref{eq:amplitude}, and it is large at weak disorder allowing $w$ to play the role of $\hbar$ from standard semiclassics~\cite{CallanColeman1977, Coleman1985}.
Higher-order corrections appear in powers of $w^2$ as we show below (with the quadratic fluctuations being $w^0$).
In this section, we will directly address whether other zero-modes (or an exponentially suppressed nearly-flat band) restore $\rho(0)=0$ via the fluctuation determinant as suggested by Refs.~\cite{BuchholdAltland2018, BuchholdAltland2018a}.
We provide a complete accounting of the relevant spectra of the fluctuations to show no such modes exist and the fluctuation determinant is $\mathcal O(1)$.

We expand $\psi = (\varphi_I + \delta\varphi,\; \chi)^T$ in flat field coordinates in order to preserve the flat measure over fields.
The superdensity decomposes with no cross terms between bosons and fermions, 
\begin{equation}
  n = \underbrace{|\varphi_I|^2}_{n_I}
    + \underbrace{\bar\varphi_I\,\delta\varphi + \delta\bar\varphi\,\varphi_I}_{\delta n_1}
    + \underbrace{\delta\bar\varphi\,\delta\varphi}_{\delta n_2}
    + \underbrace{\bar\chi\chi}_{n_F},
  \label{eq:dnsplit}
\end{equation}
and expanding the quartic in \cref{eq:E} to quadratic order decouples the two sectors. 
The last two densities combine into the manifestly graded bilinear $\delta n_{\mathrm{II}} \equiv \delta n_2 + n_F = \delta\bar\psi\,\delta\psi$.
Because \cref{eq:E} is quartic in the fields, the fluctuation action can be written out in full, and the saddle-point equations [\cref{eq:saddle,eq:amplitude}, using $w^2A^2 = \gamma$] provide the needed simplification to cancel the terms linear in fluctuations.
What remains, in the notation of \cref{eq:innerprod}, is
\begin{equation}
\begin{aligned}
  E = \hspace{-5pt}\underbrace{S_I}_{O(w^{-2})}\hspace{-5pt}
    &+ \underbrace{\int_x \big[\, \delta\bar\varphi\, K\, \delta\varphi
       + \bar\chi\, K\, \chi \,\big]
     - \frac{w^2}{2} \braket{\delta n_1, \bC * \delta n_1}}_{O(w^0)} \\
    &- \underbrace{w^2 \braket{\delta n_1, \bC * \delta n_{\mathrm{II}}}}_{O(w^1)}
    - \underbrace{\frac{w^2}{2} \braket{\delta n_{\mathrm{II}}, \bC * \delta n_{\mathrm{II}}}}_{O(w^2)},
\end{aligned}
\label{eq:Eexpand}
\end{equation}
with $S_I = s^*/w^2$ the instanton action of \cref{eq:sstar} and every order in $w$ labeled.
The first line is the classical action and fluctuations while the second represents the remaining interaction: cubic [$O(w^1)$] and quartic [$O(w^2)$].

Now consider in isolation the quadratic action, which physically represents the fluctuations. 
The self-consistent potential $U$ acts on $\delta n_{\mathrm{II}}$, which already contains both fermionic and bosonic fluctuation densities.
The same well $U$ that stabilizes the instanton also provides the potential for bosonic and fermionic fluctuations; the end result is that the kinetic and self-consistent potential pieces combine into the \emph{same} operator $K$ from before, only now $U$ is fixed and applied to both sectors.
Only $\delta n_1$ accounts for the spontaneous breaking of supersymmetry, being linear in $\delta\varphi$ with no fermionic counterpart, so $\braket{\delta n_1, \bC * \delta n_1}$ is a purely bosonic anomalous term, creating $\delta \varphi^2$ and $\delta \bar \varphi^2$ terms in the action similar to the fluctuations on top of a Bose-Einstein condensate.
The bosonic action is thus the saddle-point operator $K$ with an additional term due to the anomalous piece.
On the other hand, the fermionic quadratic action is governed only by the saddle-point operator $K$, and is therefore exactly $\int_x\bar\chi K \chi$ (\cref{sec:fermions}).

The natural way to handle the bosonic action is in Nambu space, with spinors $\hat\Phi = (\ph, \bar{\ph})^T$ and $\delta\Phi = (\delta\varphi, \delta\bar\varphi)^T$ where $\delta \bar \varphi = \delta\varphi^\dagger$; the bosonic fluctuation operator becomes
\begin{equation}
  H_B = (K \oplus \bar K) + \mathfrak{A}, \quad
  \mathfrak{A}(x,y) = -\gamma\, \bC(x-y)\, \hat\Phi(x)
  \hat\Phi^\dagger(y),
  \label{eq:HB}
\end{equation}
where $\bar K = K^*$ and the fluctuation action is $E^{(2)} = \tfrac12 \int \delta\Phi^\dagger H_B\, \delta\Phi$.
We use $\tau_{1,2,3}$ for the Pauli matrices acting on the Nambu space.
The term that mixes Nambu sectors acts on fluctuations by projecting them onto the instanton wave function pointwise and then smearing the result with the correlator $\bC$; symbolically, $\mathfrak A\,\delta\Phi = -\gamma\,\hat\Phi\;\, \bC * (\hat\Phi^\dagger \delta\Phi)$.
This object $\mathfrak A$ is unique to the bosonic sector and is ultimately responsible for reducing the fluctuation determinant to a Fredholm determinant (\cref{sec:fluctdet}).
Importantly, $H_B$ carries \emph{no dependence on $w$}: the amplitude $A$ has been scaled out, and all $w$ dependence is in the zero-mode sector.

The second line of \cref{eq:Eexpand} can be included perturbatively in $w$ with one very notable exception, the fermionic zero modes.
As we discuss in \cref{sec:zeromode} in detail, these modes must be saturated for their Grassmann integrals to be nonzero, and they \emph{only} appear in the interaction terms.
In the present section, however, everything we discuss needs only the quadratic action.

The rest of the section computes the Gaussian integral. 
\Cref{sec:bosons} identifies the bosonic zero modes and trades each for its collective coordinate, which can be safely integrated out. 
\Cref{sec:fermions} proves that the fermionic kernel is exactly the Kramers doublet (but defers the saturation of that doublet to later).
Last, \Cref{sec:fluctdet} computes the reduced determinant over the remaining non-zero modes.

\subsection{Bosonic zero modes}
\label{sec:bosons}

Every broken symmetry in the problem comes with a corresponding zero mode. 
The action for the disordered average problem is invariant under translations, rotations, and a global $U(1)$ phase, but the hedgehog solutions break all of these except for rotation about the $z$-axis combined with a phase.
With seven generators (three translations, three rotations, one phase) and one preserved, we have \emph{six} broken generators and hence six exact zero modes of $H_B$.
The resulting bosonic zero modes come in two classes, distinguished by which part of \cref{eq:HB} does the annihilating.
The phase mode $\Phi_{\mathrm{ph}} = A(i\ph, -i\bar\ph)^T$ satisfies $\hat\Phi^\dagger \Phi_{\mathrm{ph}} = iA\hat n - iA\hat n = 0$ pointwise \emph{and} $(K\oplus\bar K)\Phi_{\mathrm{ph}} = 0$.
Both terms of \cref{eq:HB} annihilate it separately. 
$J_z$ acts identically to the phase mode, thereby not contributing a new mode on its own.
On the other hand, the two transverse rotations contribute two new modes while behaving in a similar manner to the phase mode since all they do is flip the hedgehog $J_{x,y}\ph \propto \ph_{-1/2}$. 
The pointwise overlap of the two hedgehog states vanishes identically, $\ph_{+1/2}^\dagger\ph_{-1/2} = -(f^2 + g^2)\,\chi^\dagger_\uparrow \chi_\downarrow = 0$, causing a significant simplification since $\mathfrak A$ applies $\hat \Phi^\dagger(\cdot)$ before any integral is taken. 
This pointwise orthogonality ensures $\mathfrak A\, J_{x,y} \ph = 0$. 
These zero modes are the only three internal Goldstone modes, which we define as $\Phi_1$ for $x$ rotations, $\Phi_2$ for $y$ rotations, and $\Phi_3$ for phase.
Concretely,
\begin{equation}
  \label{eq:internal-boson-zeros}
  \begin{aligned}
    \Phi_1& =iA(\ph_{-1/2},\bar\ph_{-1/2}), \\  \Phi_2 & = A(\ph_{-1/2},-\bar\ph_{-1/2}), \\  \Phi_3 & =\Phi_\mathrm{ph}.
\end{aligned}
\end{equation}

Picking up translation zero modes requires the combination of $\mathfrak A$ and $K$. 
The modes are $T_a = A \partial_a \hat \Phi$, so we differentiate \cref{eq:saddle} to obtain $K\partial_a\ph = -(\partial_a U)\ph \ne 0$, and this is exactly balanced by $\mathfrak{A}\,\partial_a\hat\Phi = +(\partial_a U) \hat\Phi$, so that
\begin{equation}
  \big(K \oplus \bar K + \mathfrak{A}\big)\, T_a = 0.
  \label{eq:goldstone}
\end{equation}
In this way, $\mathfrak{A}$ encodes translation invariance from the original action; neither term annihilates $T_a$ on its own.
In other words, $H_B$ is the second variation of a translation-invariant functional, so its anomalous block \emph{must} be exactly the compensation in \cref{eq:goldstone}.

We end with a mode that, while \emph{not} a zero mode, will play a special role, the norm mode $\Phi_N = (\ph,\bar\ph)^T$.
It is the single negative mode on what will become the collective-coordinate block in \cref{sec:zeromode}, $\braket{\Phi_N,H_B \Phi_N}/\braket{\Phi_N,\Phi_N} = -(4s^*\!/\gamma)$ (which includes rotations, phases, and super-rotations of the hedgehog). 
Due to this, it helps complete the bosonic subspace related to the fermionic zeros (\cref{sec:fermions}), and it determines the steepest-descent contour (along with a half-thimble factor of $1/2$). 
Its elimination from the fluctuation determinant is also crucial to reducing it to a finite Fredholm determinant (\cref{app:normtrans}).
And finally, it appears in the collective coordinate $\Zc$, which helps us saturate the fermionic zero modes (\cref{sec:zeromode}).

The special role of the norm mode is related to its proximity to zero-modes.
At $j = \tfrac12$ the kernel of $K \oplus \bar K$ \emph{alone} is four-dimensional (four real dimensions) --- the doubled span of the two fermionic zero modes of \cref{sec:fermions}.
Three of these represent zero modes we have already identified in the bosonic sector, the phase and the two rotations; the fourth is exactly the norm mode, $\Phi_N$, for which $\hat\Phi^\dagger\Phi_N = 2\hat n \ne 0$ pointwise, so the anomalous block $\mathfrak A$ does not annihilate it.
Restricted to this four-dimensional block, we can use the above computed overlap to determine $H_B$ has eigenvalues $\{0,0,0,-4s^*\!/\gamma\}$ with $-4s^*\!/\gamma = -2.1183$. 
(On the full $H_B$, the norm mode is not an eigenstate and it mixes with other nonzero modes. \Cref{app:normtrans} tracks that coupling.)
Importantly, the norm mode is \emph{stiff} (its fractional susceptibility vanishes with $w^2$) and has steepest descent in the imaginary direction (presaged by its negative eigenvalue).

\subsubsection{Trading bosonic zeros}

Since the path integral does not depend on the exact zeroes that come about by breaking a symmetry, we can ``trade'' them for integration over the collective coordinate manifold defined by those symmetries.
This process involves taking overlaps of the zero modes to construct the Gram matrix $\mathrm{G}$, and using that as the Jacobian which accompanies the measure over the manifold of the collective coordinates.
With the manifold and its measure in hand, we can then perform its integral to obtain the volume of the collective-coordinate manifold.

Each real bosonic integral, representing a zero mode or otherwise, begins with a flat measure fixed in \cref{eq:anchor} with $d\vartheta/\sqrt{2\pi}$ (factors chosen so that the free theory integrates to one).
Along a symmetry direction, if there is just one unit-norm coordinate, it relates to the collective coordinate $c$ by $d\vartheta = \sqrt{N_c}\, dc$ with $N_c = \braket{\Phi_c,\Phi_c}$, but in the case of six flat directions the Gram matrix plays the role of Jacobian
\begin{equation}
  \prod_{c=1}^{6} \frac{d\vartheta_c}{\sqrt{2\pi}}
  \;\longrightarrow\;
  (2\pi)^{-3}\, \sqrt{\det\mathrm G}\; d^6c .
  \label{eq:trademeasure}
\end{equation}
The nonzero entries are the translation overlaps $\braket{T_a, T_b} = 2A^2 n_t \delta_{ab}$ with $n_t \equiv \int_x |\partial_z \ph|^2 = 0.6723132(24)$ [numerical], the rotation and phase overlaps $\braket{\Phi_a,\Phi_b}=2A^2 \delta_{ab}$, and the cross terms $\braket{T_a, \Phi_b} = -\frac{4s^*}{3\gamma} A^2\delta_{ab}$.
The phase and rotation tangents are not orthogonal to $T_z$ and $T_{x,y}$, respectively.
Focusing on the phase $\Phi_3$ and $T_z$ non-orthogonality, we can use the saddle equation along with projection onto the $j=1/2$ doublet to obtain the exact overlap $\braket{\ph, \partial_z \ph} = 2is^*/3\gamma$.
By symmetry, we can obtain the same overlap for $\Phi_1$ with $T_x$ and $\Phi_2$ with $T_y$. 
$\mathrm G$ is thus the direct sum of three identical $2\times 2$ blocks. 
Each block then contributes,
\begin{equation}
  \det{}_{2\times2}
  = (2A^2)\,(2A^2 n_t)\,(1-\nchi^2),
  \;
  \nchi^2 = \frac{(2s^*\!/3\gamma)^2}{n_t} = 0.185.
  \label{eq:schur}
\end{equation}
Therefore, the Jacobian is $\sqrt{\det\mathrm G} = 8 A^6 n_t^{3/2}(1-\nchi^2)^{3/2}$.
With the measure in hand, we can then exactly perform the integrals over this manifold.
Translations supply the volume $L^3$. 
Rotation and phase integrals follow the $U(1) \times SU(2)$ group acting on the $j=1/2$ doublet,
\begin{equation}
\varphi(c) = A( c_+ \ph_{+1/2} + c_- \ph_{-1/2}), \quad \lvert c \rvert = 1,
\label{eq:orbit}
\end{equation}
which through $c$ defines $S^3$ (covered once) with volume $2\pi^2$.
Along with the $(2\pi)^{-1/2}$ per bosonic zero mode, we have the zero-mode factor after performing their integrations
\begin{equation}
  B_{0} \equiv \frac{2}{\pi} A^6 n_t^{3/2} (1-\nchi^2)^{3/2}
  \label{eq:boson_zeroes}
\end{equation}
(the numerical factor coming from $(2\pi)^{-3} \times 2\pi^2 \times 8 = 2/\pi$).
We can begin to see how the prefactor depends on the disorder strength $W$; we have six factors of $A\propto 1/w$, one per bosonic zero (Coleman's $\sqrt{S_I}$ per collective coordinate~\cite{CallanColeman1977, Coleman1985}).
These six powers lead to a naive $w^{-6}$ prefactor, which we analyze in \cref{sec:powercounting}.

\subsection{Fermionic zero modes}
\label{sec:fermions}

\begin{figure}
  \includegraphics[width=\columnwidth]{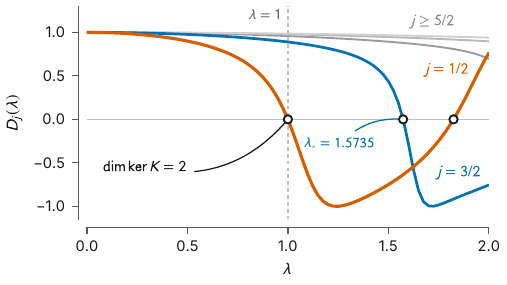}
  \caption{\textbf{Normalized radial matching determinant scan of the fermion kernel.}
$D_j(\lambda)$ for each channel $j$ as the coupling is scaled by $\lambda$; roots (circles) mark kernel elements.
At the physical coupling $\lambda = 1$ only the $j = \tfrac12$ doublet is present, $\dim\ker K = 2$; the nearest additional root sits at $\lambda_* = 1.5735$, and channels $j \ge \tfrac32$ are also excluded rigorously by the Pr\"ufer bound.}
  \label{fig:evans}
\end{figure}

The fermionic fluctuation action gives exactly the saddle operator $K$ of \cref{eq:saddle}.
The zero modes are therefore the supersymmetric doublet $\ph_{\pm 1/2}$, the two hedgehogs related by flipping $m$.
The angular momentum channel decomposition reduces each total angular momentum $j$ to a radial problem, and a Pr\"ufer bound $j+\tfrac12 > \sup_r r |U|=1.727$ (derived below) excludes every $j\ge \tfrac32$ rigorously (i.e., no other zero modes exist).
Therefore,
\begin{equation}
  \dim \ker K = 2,
  \label{eq:kerK}
\end{equation}
confirmed with a normalized radial matching determinant scan of the remaining channels (\cref{fig:evans}), which identifies the nearest additional kernel element as occurring only when the coupling is scaled, $\gamma\mapsto 1.57\gamma$ (i.e., $\lambda=1.57$ in the figure).
On the doubled (Nambu) space used below, this doublet counts as four fermionic zero modes (two complex zeros = four real zeros).

We use the rest of this section to prove this assertion, first by matching regular to decaying solutions and second via a general Pr\"ufer-angle argument.

The self-consistent potential $U = U(r)$ is spherically symmetric; thus $K$ commutes with $\bm J = \bm L + \tfrac12\bm\sigma$ and the problem separates into the radial part with the spinor spherical harmonics $\Omega_{\kappa m}$ with $\kappa = \mp(j + \tfrac12)$.
The ansatz $\chi = u(r)\, \Omega_{\kappa m} + i\, w(r)\, \Omega_{-\kappa, m}$ is closed under $K$, allowing us to write $K\chi = 0$ as two coupled first-order differential equations,
\begin{equation}
  u' + \frac{1+\kappa}{r}\, u = U w, \qquad
  w' + \frac{1-\kappa}{r}\, w = -U u,
  \label{eq:channel}
\end{equation}
of multiplicity $2j+1$; $\kappa = -1$ with $(u,w) = (f,g)$ recovers \cref{eq:radial}.
Regularity at $r=0$ demands $u \sim r^{|\kappa| - 1}$ while square-integrability demands a decay at large $r$ with $w \sim r^{- |\kappa| - 1}$ due to the short-range nature of $U\sim r^{-4}$.
For $j = \tfrac12$ this is just the instanton tail.
A zero solution to these equations exists \emph{if and only if} the regular solution $u$ continued outward \emph{becomes} the decaying ray continued inward.
The normalized radial matching determinant $D_j(\lambda)$ measures if these two potential solutions meet in the middle,
\begin{equation}
  D_j = \frac{\det\big[\Psi_{\mathrm{reg}},\, \Psi_{\mathrm{dec}}\big](r_m)}
             {\big|\Psi_{\mathrm{reg}}(r_m)\big|\,
              \big|\Psi_{\mathrm{dec}}(r_m)\big|}
  \;\in\; [-1, 1],
  \label{eq:evans}
\end{equation}
with $\Psi = (u,w)^T$ integrated outward ($\Psi_\mathrm{reg}$) and inward ($\Psi_\mathrm{dec}$); by Abel's identity $r^2 \det[\Psi_{\mathrm{reg}}, \Psi_{\mathrm{dec}}]$ is exactly constant, so the matching radius $r_m$ is immaterial and $D_j$ vanishes if and only if the $j$ contributes to $\ker K$.

The scan of this quantity in \cref{fig:evans} reveals when other zero modes can occur. 
Here, we hold the converged $U$ fixed and scale it so that we can find the zeros for the whole family $K_\lambda = i\bm\sigma\cdot\nabla + \lambda U$.

At $\lambda = 1$ the only root is the $j = \tfrac12$ doublet ($|D_{1/2}| = 4.5\times10^{-8}$, numerical accuracy), with $D_{3/2} = 0.889$, $D_{5/2} = 0.957$ rising monotonically with $j$ toward $1$; the nearest additional kernel event is a $j = \tfrac32$ four-fold degenerate manifold at $\lambda_* = 1.5735(1)$.
By this measure, \cref{eq:kerK} is true with a $57\%$ margin.

More generally, higher $j$ channels \emph{cannot} contribute.
In fact, we can prove the following exclusion criterion
\[
  j + \tfrac12 > \sup_r r|U(r)|
  \quad\Longrightarrow\quad
  \text{no channel $j$ zero mode}.
\]
The proof follows a simple argument: setting $u = R\cos\theta$, $w = R\sin\theta$ ($\theta$ being the Pr\"ufer angle), we obtain a simple equation for the angle derived from \cref{eq:channel}
$$\theta' = -U + (\kappa/r)\sin 2\theta.$$
Any zero solution must rotate $\theta$ from $0$ (the regular $r=0$ solution) to $\tfrac\pi2$ (the decaying $r\rightarrow \infty$ solution) through $\theta = \tfrac\pi4$.
However, at $\pi/4$, the centrifugal term maximally opposes this flow; thus the potential must overcome it $r|U(r)| \ge j + \tfrac12$ at whichever $r$ this occurs.
Consequently, we can numerically but rigorously exclude all channels $j \ge \tfrac32$ ($\sup_r r|U| = 1.727 < 2$).

Unlike the bosonic zero modes, fermionic zeroes cannot simply be integrated over since Grassmann integration demands $\int d\chi (1) = 0$.
The integrals survive only if the modes are ``saturated'' --- found explicitly somewhere in the integrand; the $\bar\psi\mathsf k\psi$ insertion provides this for one pair, but for the other pair we must look to the interaction terms.
Doing this carefully is the major subject of \cref{sec:zeromode}.
In the meantime, we look at the fluctuation determinant for all remaining modes.

\subsection{The fluctuation determinant}
\label{sec:fluctdet}

\begin{figure}
  \includegraphics[width=\columnwidth]{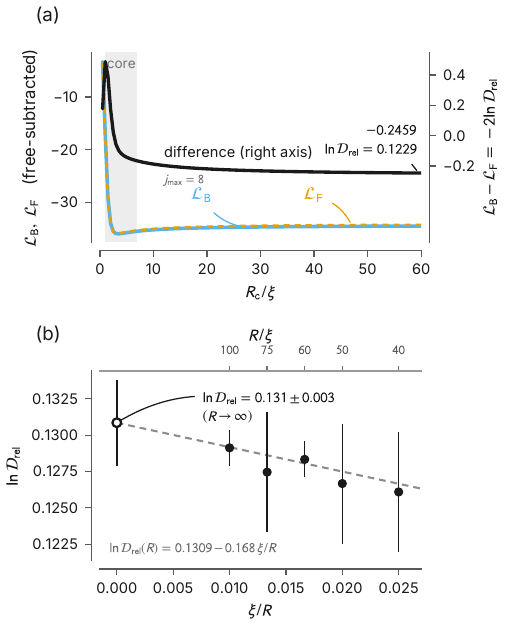}
  \caption{\textbf{The fluctuation determinant.}
(a) Radial attribution of the fluctuation weight, \cref{eq:radialattr}: each eigenvalue's logarithm is weighted in proportion to its eigenvector's norm in $r<R_c$, so the curves show the free-subtracted partial log determinants $\mathcal L_{\mathrm{B}}$, $\mathcal L_{\mathrm{F}}$ accumulated out to radius $R_c$ ($j_{\max} = 8$, $R = 60\,\xi$): each separately runs to $\approx -34$, and they cancel to the residual $-0.2459 = -2\ln\Drel$ at this truncation (right axis), exactly the supersymmetric cancellation of determinants which enables \cref{eq:anchor}.
The shaded band marks the instanton core, $1 \le r/\xi \le 7$, where the difference accumulates; beyond $R_c = 20\,\xi$ only $-0.028$ remains.
(b) $\ln\Drel$ after $j_{\max}$ extrapolation versus $\xi/R$ for $R/\xi = 40$--$100$ (top axis); the weighted linear fit gives the quoted $\ln\Drel(R\to\infty) = 0.130(3)$, the open symbol at $\xi/R = 0$.  }
  \label{fig:fluct}
\end{figure}

What remains of the Gaussian integrals is the superdeterminant (with zero modes removed). 
To understand the difference between bosonic and fermionic determinants, we first regulate both with the free Weyl operator,
\begin{equation}
  \ln \Drel = \frac12
    \ln\left|\frac{{\det}'\big[(K\oplus\bar K)/H_{\mathrm{free}}\big]}
                  {{\det}'\big[H_B/H_{\mathrm{free}}\big]}\right|,
  \label{eq:Drel}
\end{equation}
where the prime indicates the removal of \emph{only} the zero-modes we enumerated above (six bosonic and four fermionic)\footnote{We will track signs and phases in \cref{app:thimble}, concentrating only on the magnitude of things.}.
$H_{\mathrm{free}} = (i\bm\sigma\cdot\nabla)\oplus\overline{(i\bm\sigma\cdot\nabla)}$ is the free Nambu-doubled Weyl operator (i.e., $U\equiv0$).
$H_\mathrm{free}$ regulates both fluctuation determinants since unregulated both are formally infinite and dimensionful (their ratio, however, is the relevant quantity and does not need this arbitrary regularization, used here just for illustrative purposes).
Due to the division of determinants, the influence of $H_\mathrm{free}$ will cancel identically when $\Drel$ is computed.
It is useful to define the log-determinants
\begin{equation}
\begin{aligned}
  \mathcal L_{\mathrm B} &\equiv \ln\big|{\det}'\big[H_B/H_{\mathrm{free}}\big]\big|,\\
  \mathcal L_{\mathrm F} &\equiv \ln\big|{\det}'\big[(K\oplus\bar K)/H_{\mathrm{free}}\big]\big|,
\end{aligned}
\label{eq:LBLF}
\end{equation}
so that $\ln\Drel = -\tfrac12\,(\mathcal L_{\mathrm B} - \mathcal L_{\mathrm F})$.
This is done in \cref{fig:fluct} for the Gaussian correlator, a computation we describe in detail below.

Numerically, we compute \cref{eq:Drel} in a truncated free basis and extrapolate the angular cutoff and the domain radius while removing only the known zero modes by testing overlap with the exactly computed zero modes above (requiring overlap to exceed $0.999$). 
The results are shown in \cref{fig:fluct}.
To do this computation, we use the eigenbasis of $H_\mathrm{free}$ for a ball of radius $R$ with hard-wall boundary conditions and truncated at a maximum energy of $E_{\max} = 12$ and $j_{\max}$.
The instanton's spin texture (the hedgehog) enters through $\mathfrak A$ and hence it can couple different angular momenta across Nambu sectors while preserving
\begin{equation}
  \label{eq:Moperator}
  \hat M \equiv J_z\oplus \overline{ J_z} - \tfrac12 \tau_3,
\end{equation}
with $\overline{J_z} = - J_z^*$. 
The operator $\hat M$ represents a symmetry (the unbroken phase and rotational symmetry combination) and has eigenvalues labeled $M \in \mathbb Z$; the $M$ are the \emph{magnetic quantum numbers} relative to the hedgehog. 
Charge conjugation in Nambu space will send $M\rightarrow -M$; $M=0$ is the self-conjugate sector.
We can thus block-diagonalize across these $M$-sectors to compute $\mathcal L_{\mathrm B}$ and $\mathcal L_{\mathrm F}$ as sums over the eigenvalues we have kept $\sum_n \ln |\lambda_n|$ while subtracting the same sum for $H_\mathrm{free}$.
Zero modes are removed by \emph{identity}: an eigenvector is discarded only if it overlaps one of the analytically known zero modes above the $0.999$ threshold.
Removal by identity leaves room for a near-zero band of eigenvalues to emerge.
In fact, whether anomalously small eigenvalues exist is exactly the question raised by Refs.~\cite{BuchholdAltland2018, BuchholdAltland2018a}, so nothing may be dropped merely for being small.
The two truncations are then increased to converge the result.
As $j_{\max}$ increases, we observe that the increment in the result decreases geometrically (consecutive values have ratio $\approx 0.23$) such that we can extrapolate to $j_{\max}\to \infty$ (with uncertainty in extrapolation dominating).
On the other hand, the $R$ dependence is observed to be linear in $\xi/R$ over $R\in[40,100]\, \xi$ allowing simple extrapolation to the infinite-volume value [\cref{fig:fluct}(b)].

Stepping back, we can spatially resolve where the fluctuation determinant begins to differ strongly from the free theory.
\Cref{fig:fluct}(a) demonstrates how supersymmetry enables a large cancellation between bosonic and fermionic contributions within the core of the instanton; this is computed with the log of eigenvalues weighted by the integrated density of the eigenfunction within a ball of radius $R_c$.
Intuitively, each eigenvalue will be attached to the region in which its eigenmode lives.
Precisely, every eigenvector has a radial density $\rho_n(r)$ (angular momentum channels are squared and summed), and we can define the $R_c$-dependent quantity
\begin{equation}
\begin{gathered}
  \mathcal L_{\mathrm{B,F}}(R_c)
  = \sum_n \ln\big|\lambda^{(B,F)}_n\big|\, f_n(R_c)
    - \big(\text{same for } H_{\mathrm{free}}\big),\\
  f_n(R_c) = \int_{r \le R_c}\!\rho_n(r) ,
\end{gathered}
\label{eq:radialattr}
\end{equation}
At full radius $f_n(R) = 1$, so $\mathcal L_{\mathrm{B,F}}(R) = \mathcal L_{\mathrm{B,F}}$ which reproduces the log determinants \cref{eq:LBLF}.
Each sector separately begins to differ from the free problem by $\approx -34$ of log weight.
Despite this large difference, the two curves track each other nearly exactly at every radius, and the raw spectral sums (without the free-subtraction) cancel to $2.5$ parts in $10^6$, leaving behind a remnant of difference stabilizing at $-0.246$ (the unsubtracted quantities $\sum_n \ln |\lambda_n| f_n$ each reach $\approx 10^5$). 
Additionally, this difference ($-2\ln\Drel$, right axis) saturates to a large degree within the instanton core (shaded band).
The region in the instanton's tail $R_c > 20\,\xi$ contributes only $-0.028$.
We make this precise below in computing the Fredholm determinant, but in a word, the quadratic fluctuation operators between bosons and fermions essentially match away from the instanton core.
The extrapolated value is
\begin{equation}
  \ln\Drel = 0.130 \pm 0.003, \qquad \Drel \approx 1.14.
  \label{eq:Drelnum}
\end{equation}
A reader comparing the two panels of \cref{fig:fluct} against this number should note the truncation ladder: the plateau of panel (a) sits at the fixed truncation $j_{\max} = 8$, $R = 60\,\xi$, where $\ln\Drel = 0.1229$; extrapolating $j_{\max}\to\infty$ raises it to $0.1283$, which is the $R = 60\,\xi$ point of panel (b); and the $R\to\infty$ intercept gives the quoted $0.130$.
The shifts are systematic and directional, which is why panel (a)'s plateau is not $-2\times0.130$.

On the other hand, because $\mathfrak{A}$ is the only difference between the bosonic and fermionic Hessians, $\Drel$ collapses to a Fredholm determinant of a localized operator,
\begin{equation}
  \Drel^{-2} = \frac{4s^*}{\gamma} \bigg[ \frac{2n_t(1-\nchi^2)}{h_T} \bigg]^{3} {\Det}'\big(1 + G_{I}' \mathfrak{A}_\mathrm{eff}\big),
  \label{eq:fredholm}
\end{equation}
with $G_{I}'$ the doubled fermionic zero-energy Green's function with the zero-mode sector removed (the inverse of $K \oplus \bar K$ on the complement of its kernel), $h_T \equiv \gamma\braket{\partial_z \hat n, \bC * \partial_z \hat n}$, and
\begin{equation}
\mathfrak{A}_\mathrm{eff}(x,y)
= -\gamma \hat\Phi(x)[\bar C(x-y) - \tfrac{1}{2s^*}U(x)U(y)]\hat\Phi^\dagger(y),
\label{eq:Aeff}
\end{equation}
the Schur complement of the bosonic sector with the norm mode eliminated (a zero of the fermionic sector); the factor $\frac{4s^*}{\gamma}$ in front is the norm matrix-element extracted by that elimination.
The extra factor $[2n_t(1-\nchi^2)/h_T]^3$ is the kernel metric left behind by removing the three bosonic translation zero modes (which are not zero modes of the fermionic $K\oplus \bar K$).
This expression is particularly convenient compared to \cref{eq:Drel}, requiring no cutoff or division of two formally infinite expressions. 
It is mathematically a Fredholm determinant ${\Det}(1+T)$ whose formal definition is through the spectrum of $T$. 
For our case, $T = G_I'\mathfrak A_{\mathrm{eff}}$ is built out of the self-consistent Green's function and instanton profile encoded in $\mathfrak A_{\mathrm{eff}}$, resembling a $T$-matrix scattering problem.
Just like a scattering problem, the content of this determinant is dictated by the localization of $\mathfrak A$ to the instanton's core.
Details of how to obtain this expression are in \cref{app:normtrans}; the end result is that $\Det'$ has all the identified fermionic and bosonic zeroes removed, resulting in a well-behaved operator $G_{I}'\mathfrak{A}_\mathrm{eff}$\footnote{We conjecture this operator is trace class and has no extra $-1$ eigenvalues (i.e., zeroes of the bosonic fluctuation determinant) \cite{ShipmanWilson2026}.
The numerical evidence here strongly supports this claim.}.
Therefore, fluctuations are $|\ln\Drel| = \mathcal O(1)$ ($w$-independent). 
Importantly, this means that the double-exponential suppression $\exp(-2e^{S_I/3})$ proposed in Ref.~\cite{BuchholdAltland2018} would require $G_{I}'\mathfrak{A}_\mathrm{eff}$ to have eigenvalues that grow like the exponential of $\mathcal O(e^{S_I/3})$ while still remaining trace class (numerically $\|G_I' \mathfrak{A}_{\mathrm{eff}}\|_1 = 11.3$).
This can be attributed to \Cref{eq:anchor}: outside of the core of the instanton, both bosonic and fermionic sectors look approximately the same with identical $K$ and thus their contributions cancel one-for-one, leaving only the contribution from $\mathfrak{A}$.

Numerically evaluating the Fredholm determinant \cref{eq:fredholm} allows for a technically different way to compute $\Drel$ using the spectrum of $G_{I}'\mathfrak{A}_\mathrm{eff}$.
At fixed truncation ($E_{\max}$ and $j_{\max}$), this matches the separate bosonic and fermionic sector computation $\mathcal L_B - \mathcal L_F$ of \cref{fig:fluct} to $10^{-11}$ (not a statement of accuracy, just of consistency of the two approaches).
The same spectrum locates the eigenvalue $-1$ directions of $G_I'\mathfrak{A}_\mathrm{eff}$: precisely the three translation Goldstones (measured at $\mu = -1.00033$), and nothing else.
(The value $-1$ is forced: a translation is a zero mode of $H_B$ but not of $K\oplus\bar K$, so in the Fredholm form it must annihilate a factor $1+\mu$; the primed $\Det'$ omits exactly these three factors, whose contribution is to the $[2n_t(1-\nchi^2)/h_T]^3$ term of \cref{eq:fredholm}.
Note that we find no other $\mu$ near $-1$, which tells us that there are no other bosonic zeros.)
Any exponential domain dependence of $\ln\Drel$ is excluded directly at the level $3\times 10^{-5}/\xi$ over $R \in [40, 100]\,\xi$.
The chain of bounds sits far below its ceiling: at $R = 60\,\xi$ the chain runs $|\ln{\Det}'| = 2.32 \le \sum_j|\mu_j| = 5.13 \le \|G_I'\mathfrak A_{\mathrm{eff}}\|_1 = 11.3$ (\cref{app:normtrans}).

\section{The zero-mode sector and the prefactor}
\label{sec:zeromode}

With the full accounting of the fluctuation determinant, the only thing left is the fermionic zero-mode sector and the density-of-states insertion $\bar \psi \mathsf k \psi$ which we can handle all at once and thereby compute the total power of $w$ of the prefactor.
We have already accounted for $w^{-6}$ via the bosonic sector and the six zero modes that were traded for six factors of the instanton amplitude (\cref{sec:bosons}).
The density-of-states insertion supplies one more factor of $A^2$ on the condensate, and the fermionic zero modes appear to be saturated by the interaction terms [$O(w^2)$] balancing the $A^2$ from the density-of-states insertion; this is Coleman's counting~\cite{Coleman1985}, and it gives the prefactor $\sim w^{-6}$.
This answer is wrong; its coefficient vanishes identically due to supersymmetry.
Since the instanton is entirely in the bosonic sector, supersymmetry appears to be spontaneously broken but is restored via the zero-mode sector and encoded in a set of exact Ward identities whose practical significance is to annihilate the lowest order term leading to $w^{-6}$; it appears as Parisi--Sourlas dimensional reduction that guarantees $Z=1$.
We therefore get a nonzero result at the next order, including one relative-fluctuation propagator $1/S_I = w^2/s^*$ (in other words, ``one-loop''), leading directly to the prefactor $\sim w^{-4}$.
In this section, we show how this occurs and compute the coefficient. 

\Cref{sec:powercounting} follows through with the naive $w^{-6}$ argument to show both how it fails and how to fix it. 
\Cref{sec:supermatrix} introduces the collective coordinate as a supermatrix and with it we derive the Ward identities. 
\Cref{sec:frames} discusses the body and covariant frames for the collective coordinates while \cref{sec:wardreduction} performs the power-counting argument precisely using Ward identities to saturate the fermionic zero modes.
\Cref{sec:kappa} disposes of the single form factor that could have modified the result. 
And finally, \cref{sec:assembly} combines everything into the final result \cref{eq:result}.

\subsection{Why the naive $w^{-6}$ has a vanishing coefficient}
\label{sec:powercounting}

Before carefully considering the saturation of fermionic modes in the context of Ward identities in \cref{sec:supermatrix}, it is useful to see how it comes about by finding the terms in the action which contain the fermionic zeroes.
We give an argument which naively gives the wrong power of $w$ if we do not carefully consider cancellation between two terms. 
This cancellation will be related to the contact identity derived in \cref{sec:supermatrix}.

The bosonic trading of zero modes for integrations in \cref{eq:boson_zeroes} carries one bare factor of $A$ per traded mode, so $B_0 \propto A^6 = \gamma^3/w^6$.
The density-of-states insertion $\bar\psi\mathsf k\psi$ has its lowest order contribution as that of the condensate amplitude, $\int_x n_I = A^2$, contributing $w^{-2}$.
The fermionic zero modes exist in the interacting terms; all four together can be found by expanding the quartic term in \cref{eq:Eexpand}
\begin{equation}
  \frac{w^2}{2} \int_{x,y} (\bar\chi\chi)_x\, \bC(x-y)\,
  (\bar\chi\chi)_y.
  \label{eq:fquartic}
\end{equation}
There is one more term at this order to consider: the cubic boson-fermion interaction in \cref{eq:Eexpand} that goes as $O(w^1)$.
We defer that term for a moment (it is the term that supersymmetry demands must balance the quartic term).
The quartic \cref{eq:fquartic} appears in the weight $e^{-E}$ with a \emph{plus} sign by the contour rotation of \cref{sec:model}.
Writing $\chi = \alpha_+\ph_{+1/2} + \alpha_-\ph_{-1/2} + (\mathrm{massive})$ and recalling the pointwise doublet orthogonality from \cref{sec:saddle}, we have $\bar\chi\chi = \hat n\,(\bar\alpha_+ \alpha_+ + \bar\alpha_- \alpha_-) + \cdots$ with no cross terms. 
The expanded quartic term then saturates the Grassmann integrals at first order ($\int d\bar\alpha_+ d\alpha_+ d\bar\alpha_- d\alpha_-\; \bar\alpha_+ \alpha_+ \bar\alpha_- \alpha_- = 1$) producing
\begin{equation}
  Q_4 \equiv w^2\braket{\hat n, \bC \ast \hat n} = \frac{2s^*w^2}{\gamma^2}.
  \label{eq:Q4}
\end{equation}
The product $A^2 \cdot Q_4 = O(w^0)$, contributing nothing to the powers already counted by $B_0$, so that assembled factors give $A^6\,(2s^*\!/\gamma) = 2s^*\gamma^2 w^{-6}$, the Coleman counting~\cite{Coleman1985} for six bosonic collective coordinates and a noncontributing fermion sector.
This appears to yield $\rho(0) = \mathcal A_{\mathrm{naive}}\, w^{-6} e^{-s^*/w^2}$.

However, $\mathcal A_\mathrm{naive}$ is identically zero once we consider the neglected term as well as a measure term developed more fully in \cref{sec:supermatrix}.
The missing measure term will be that of the norm coordinate $n_I \mapsto p\, n_I$ where $p$ is allowed to fluctuate.
As we show in \cref{sec:supermatrix}, the insertion splits as $\bar\psi\mathsf k\psi = \bar\psi\psi - 2\bar\chi\chi$ with its bosonic piece living entirely in the supertrace and therefore vanishing identically via \cref{eq:contact}.
To see the cancellation, consider how the norm mode $\delta = p - 1$ couples quadratically to the fermionic zero modes in the $O(w^1)$ term. 
(This is an abbreviated accounting; a full accounting requires the norm mode's coupling within the quartic term as well, which only makes this argument more robust away from the saddle point.) 
The term in $E$ this corresponds to is
\begin{equation}
   -\gamma \delta \braket{\hat n, \bC*\hat n} (\bar \alpha_+ \alpha_+ + \bar \alpha_- \alpha_-).
   \label{eq:zero-mode-mass}
\end{equation}
Therefore $\mathfrak m(\delta) \equiv -\gamma \delta \braket{\hat n, \bC*\hat n}$ is the fluctuating mass of the zero modes.
The descending contour allows us to compute $\braket{\delta^2} = -1/2S_I$ (see \cref{app:thimble}), and therefore 
\begin{equation}
  \braket{\mathfrak m^2} + Q_4
  = \big(\gamma\braket{\hat n,\bC*\hat n}\big)^2\Big({-}\frac{1}{2S_I}\Big) + w^2\braket{\hat n,\bC*\hat n} = 0,
  \label{eq:susycancel}
\end{equation}
identically.

All we are left with is the fermion bilinear $-2\bar \chi \chi$.
Since it already saturates one complex zero mode on its own, the partner pair cannot come from the quartic term alone (by nilpotency, both complex zero modes are required for this term), and therefore comes from $\mathfrak m(\delta)$.
However, $\mathfrak m(\delta)$ is linear in $\delta$, so to obtain a nonzero value, we need the contribution of the \emph{measure} of \cref{sec:supermatrix}, where $p \, dp \mapsto (1+\delta) \, d\delta$ and hence we obtain the factor $\braket{(1+\delta)\delta} = -1/2S_I$. 
With no balancing $A^2$ from the insertion, the surviving term is lowered by $2/S_I$.
Physically, $1/S_I = w^2/s^*$ because the instanton is \emph{stiff}: at weak disorder its amplitude barely fluctuates, and that stiffness is the origin of the surviving $w^2$.
The complete counting of negative powers is therefore $A^6 \sim w^{-6}$ from the Jacobians and the relative reduction by $2/S_I \sim w^2$:
\begin{equation}
  \rho(0) \sim w^{-6}\cdot w^{2} e^{-S_I} = w^{-4} e^{-S_I}.
  \label{eq:powershift}
\end{equation}

This method has potential flaws. For instance, did we find all of the terms and collect them appropriately? 
We used the fluctuating mass and Jacobian to fix things and argued for excluding the quartic term.
Furthermore, does this argument survive the zero modes coupling to the other terms?  
To resolve these issues, the rest of this section shows how this works \emph{exactly}.

\subsection{The instanton as a supermatrix}
\label{sec:supermatrix}

Before we resolve the issue of fermionic zero modes, we need to address something that was previously swept under the rug: Does the instanton solution violate supersymmetry?
More precisely, can we define a collective coordinate for the instanton such that \cref{eq:anchor} is satisfied?
We can verify this with elementary algebra before moving to the effects of the fluctuation determinant, and in doing so, we set the stage for everything that follows.
In the end, we find that supersymmetry survives via the zero-mode sector as a set of exact Ward identities, and those identities will be crucial for saturating the Grassmann zero-mode integrals.

The crucial part of the supersymmetric approach is how \cref{eq:anchor} has the same differential operator for bosonic and fermionic fields.
However, as we have seen, the instanton solution broke not only supersymmetry but $J_z$ spontaneously; as such, any collective coordinate needs to include the full Kramers doublet $\ph_m$ of \cref{sec:saddle} as well as bosonic and fermionic realizations.
As a result, the instanton can be parameterized such that
\begin{equation}
  \Psi_0 = A\,\sum_m \ph_m(x - X)
  \begin{pmatrix} z_m \\ \zeta_m \end{pmatrix},
  \quad
  \Zc =
  \begin{pmatrix} z_+ & z_- \\ \zeta_+ & \zeta_- \end{pmatrix},
  \label{eq:Zdef}
\end{equation}
with $z_m$ the complex bosonic amplitudes and $\zeta_m$ Grassmann amplitudes collected into the $(1|1)\times2$ supermatrix $\Zc$, and $X$ the instanton position.
Note what this has done naturally: all bosonic and fermionic zero-modes have been accounted for but in the process we have mixed in one massive mode, the norm mode.
A few combinations of these quantities will be convenient to define, including $p = z^\dagger z$, $N_\pm = \bar\zeta_\pm\zeta_\pm$, and $\nu = \bar\zeta\zeta = N_+ + N_-$.

Recall the pointwise orthogonality $\ph_+^\dagger(x)\ph_{-}(x) = 0$, which helps us compute a number of quantities without issue.
For instance, the superdensity with the collective-coordinate \cref{eq:Zdef} is $n(x) = A^2\hat n(x)\,q$ with the \emph{single} superinvariant
\begin{equation}
  q = p + \nu.
  \label{eq:qdef}
\end{equation}
With that invariant, the action \cref{eq:E} evaluates \emph{exactly} under $\Psi_0$ to
\begin{equation}
  E = S_I\,(2q - q^2),
  \label{eq:Eq}
\end{equation}
with $S_I = s^*/w^2$ as in \cref{eq:sstar}.
The supermatrix $\Zc$ helps us understand the invariants in the problem.
Act on the right by a $U(2)$ rotation and one chooses a different direction for the hedgehog; act on the left by a super-rotation and one moves the instanton to a mixture of bosonic and fermionic sectors. 
Invariance under these operations is captured by the invariants of $\mathsf R = \Zc^\ddagger\Zc$ (with $\ddagger$ the graded adjoint), namely, $q = \tr\mathsf R$ and the nilpotent
\begin{equation}
  d = \det\mathsf R , \qquad d^2 = 0.
  \label{eq:ddef}
\end{equation}
Here we define the \emph{body frame} as
\begin{equation}
  z_+^\mathrm{body} = \sqrt{p}, \quad z_-^\mathrm{body} = 0.
  \label{eq:bodygauge}
\end{equation}
(We will drop the ``body'' superscript where there is no ambiguity.)
In the body frame the nilpotency is explicit, $d = p N_- + 2N_+N_-$.
This parametrization allows us to separate out the zero-modes from the fluctuations in one go and gives us precise statements about the form of the result.
We begin by writing $\psi = \Psi_0[\Zc] + \Xi$ with the massive modes in $\Xi$ which are integrated out at a fixed $\Zc$ leading to
\begin{equation}
  \mathcal H(\Zc)\, e^{-E(q)} \equiv \int \mathcal D[\bar\Xi,\Xi]\;
    e^{-E[\Psi_0[\Zc] + \Xi]},
  \label{eq:Hdef}
\end{equation}
where the measure still includes the translation modes.
In \cref{sec:bosons}, we deal with the zero-mode translations, relating them to integrals over the corresponding symmetry manifolds. 
Those integrals and their Jacobian factors, in the collective-coordinate language, can (and do) depend on $\Zc$ as we deal with below in \cref{sec:frames,sec:wardreduction}.
Because these objects must be invariant under both the above group actions, $\mathcal H$ can depend on $\Zc$ only through $q$ and $d$.
As such, we must have 
 $\mathcal H(\Zc) = H_0(q) + d\,\varkappa(q)$ (linear in $d$ due to its nilpotency). 
 $H_0$ and $\varkappa$ remain arbitrary for the moment; we compute them in \cref{sec:wardreduction}.

We can now state the Ward identities in terms of Berezin calculus.
With the Gaussian normalization implicit in \cref{eq:anchor} (one factor of $1/\pi$ per complex boson), the collective-coordinate measure is flat and reduces to $d^4z\,d^4\zeta/\pi^2 \mapsto p\,dp\,d^4\zeta$ after integrating out the $S^3$ rotation symmetry directions.
The remaining Grassmann integrals in need of saturation follow for any smooth $\mathcal W$,
\begin{equation}
\begin{gathered}
  \int d^4\zeta\, \mathcal W(q) = \mathcal W''(p), \qquad
  \int d^4\zeta\, \nu \mathcal W(q) = 2\mathcal W'(p), \\
  \int d^4\zeta\, d\, \mathcal W(q) = p\mathcal W'(p) + 2\mathcal W(p).
\end{gathered}
\label{eq:berezin}
\end{equation}

It is worth pointing out two facts. 
First, we will now \emph{not} need the angular Jacobian since that has been accounted for via the integral over $S^3$ and the $\pi^{-2}$ from the normalization of the bosonic integrals.
Second, the inclusion of $A$ in \cref{eq:Zdef} is invisible to the measure as a consequence of changing variables for both bosonic and fermionic variables (the \emph{Berezinian} being the generalized Jacobian for this mixed measure).
Explicitly, in terms of the original flat coordinates $a_m = Az_m$ and $\alpha_m = A\zeta_m$, we have the measures $d^4z = A^{-4}d^4a$ and $d^4\zeta = A^{+4}d^4\alpha$.
Clearly, the product carries \emph{no} power of $A$ at all,
\begin{equation}
  d^4z\,d^4\zeta = d^4a\,d^4\alpha,
  \qquad \Ber\big(A\,\mathds 1_{4|4}\big) = A^4/A^4 = 1 .
  \label{eq:Aneutral}
\end{equation}
This will lead to a subtle difference from the bosonic $B_0$ computed earlier in that the powers of $A$ due to rotations do not appear as a result of rescaling the Grassmann variables as well, and we will only get three powers of $A$ due to the translation modes (as we compute in \cref{eq:JX}).

We can now check the normalization of our partition function; that is, we must ensure $Z=1$.
The partition function with our reduction and collective-coordinates must be
\begin{equation}
  Z \equiv \int \mathcal D[\bar\psi,\psi]\, e^{-E[\psi]}
  = \int_{\mathcal J} dp\; p \int d^4\zeta\;
    \mathcal H(\Zc)\, e^{-E(q)}.
  \label{eq:Zfull}
\end{equation}
Here $\mathcal J$ is the curve the norm coordinate $p$ traverses, and we now need to be careful about the $e^{i\pi/4}$ rotation of \cref{sec:model}.
The rotation mapped the original norm direction on the half-line onto $\mathcal C_0 = -i[0,\infty)$. 
We can check this by noting that $\mathrm{Re}\,E = S_I(\mathrm{Im}\,p)^2 \to +\infty$ ensuring convergence of the integral.
The original curve $\mathcal J$ may be taken as $\mathcal C_0$ itself or as any analytic deformation holding its two ends fixed ($p=0$ and the $\mathrm{Im}\, p \rightarrow -\infty$ region).
The convenient deformation for us is the continuation through the saddle at $p_* = 1$ constructed in \cref{app:thimble}.
Writing $F(t) = H_0(t)e^{-E(t)}$ and $\mathcal W_\varkappa(t) = \varkappa(t)e^{-E(t)}$, the identities \cref{eq:berezin} reduce the integrand of \cref{eq:Zfull} to a total derivative and since $e^{-E}\to 0$ at one end
\begin{equation}
  Z = \Big[\, pF' - F + p^2 \mathcal W_\varkappa \,\Big]_{\partial\mathcal J} = H_0(0).
  \label{eq:noinsertion}
\end{equation}
No saddle analysis is necessary; the endpoints suffice to fix $Z$.
At $\Zc = 0$ there is no instanton, and the remaining transverse integral in \cref{eq:Hdef} is purely the normalization integral \cref{eq:anchor} and therefore $H_0(0) = 1$, ensuring that 
\begin{equation}
  Z = 1,
  \label{eq:Zone}
\end{equation}
exactly, on $\mathcal C_0$ or on any deformation of it.
This is a form of Parisi--Sourlas dimensional reduction \cite{ParisiSourlas1979,ParisiSourlas1982} implying that the instanton cannot contribute to $Z$.

The same algebra with $\mathcal W \to t\mathcal W(t)$ gives the second identity,
\begin{equation}
  \int dp\; p \int d^4\zeta\; q\,[H_0 + d\varkappa]e^{-E} = 0,
  \label{eq:contact}
\end{equation}
the \emph{contact identity}.
With an extra power of $p$ after the final integral, the term at the $p=0$ end of $\mathcal J$ also vanishes regardless of whether we pass through the instanton or not.  
This is to be expected from the fact that the supertrace $\bar \psi \psi \mapsto A^2 q$ must vanish, consistent with the vanishing bilinear noted below \cref{eq:insertion}.

\Cref{eq:contact} plays a role in the power-counting argument of \cref{sec:powercounting}.
To understand this, the density-of-states insertion can be written as a supertrace and subtraction of the Grassmann variables $\bar\psi\mathsf k\psi = \bar\psi\psi - 2\bar\chi\chi$.
The supertrace part corresponds to what cancelled identically above, and all that is left is the fermionic insertion $\bar \chi \chi$.
(Due to this, we could have begun with $\bar \chi \chi$ leading to the Green's function $G^+_{\omega,x,x'}$.)

\subsection{Body frame and covariant frame}
\label{sec:frames}

With the partition function confirmed to integrate to unity, we will need to compute the Green's function next.
There is a danger that we may violate a Ward identity (such as \cref{eq:noinsertion,eq:contact}) by choosing a \emph{frame} which treats transverse fluctuations and collective coordinates inconsistently.
We head this off by consistently \emph{imposing} the Ward identities, keeping the $U(2) \times U(1|1)$ structure of \cref{sec:supermatrix} at every step.
In the process, we will saturate the fermionic zero modes, reduce the problem to the fluctuation determinant already computed, and show the only new term, a form factor $\varkappa_1$, vanishes identically by symmetry.

As we have already done, we split the field $\Psi = \Psi_0[\Zc, X] + \Xi$ with $\Psi_0$ encoding the zero-mode configuration \cref{eq:Zdef} and $\Xi = (\eta\,|\,\xi)$ the transverse fluctuation.
Fix the translation patch at $X = 0$.
Note that $\Xi$ differs slightly from \cref{sec:supermatrix} in that we are now explicitly removing the translation modes which we collect into a Jacobian factor and fix the configuration to be around $X=0$.
The transverse fluctuations, being orthogonal to the zero modes, will change with $\Zc$, the frame we choose.
In the body frame \cref{eq:bodygauge}, the frame our computations use, and the one used throughout below, the constraints are fixed and do not change with $\Zc$: $\eta$ is orthogonal to the bosonic kernel and translation directions, $\xi$ to $\ker K$.
In a \emph{covariant frame} the constraints instead rotate with $\Zc$, which makes the symmetry manifest at every algebraic step but is substantially more difficult in practice (and it is not what we implement numerically). 
We check only a few key aspects of the covariant frame to ensure the body frame does not miss any crucial term from the Berezinian.

Naturally, the change of variables from $\Psi$ to $(X,\Zc,\eta,\xi)$ carries a Berezinian, the superspace Jacobian of the frame matrix we call $\mathcal B$
\begin{equation}
  \mathcal B = \begin{bmatrix} \mathcal B_{bb} & \mathcal B_{bf} \\ \mathcal B_{fb} & \mathcal B_{ff} \end{bmatrix},
 \label{eq:frameM}
\end{equation}
computed with the overlaps of the respective modes corresponding to each collective coordinate.
This is particularly simple in the body frame; there is no $\Zc$ dependence in the transverse fluctuations and the mixed blocks $\mathcal B_{fb}$ and $\mathcal B_{bf}$ vanish due to 
$\partial_{\zeta_m}\Psi_0 = A\ph_m\, ( 0 | 1 )^T$ being entirely in the fermion block.
As a result, $\Ber^{\mathrm{body}} = \det \mathcal B_{bb}/\det \mathcal B_{ff}$.
The result is therefore \emph{pure body} with no Grassmann completion, no dependence on the transverse fields, and no Faddeev--Popov or connection terms:
\begin{equation}
  \Ber^{\mathrm{body}} = A^3\, p^{3/2}\, n_t^{3/2}(1 - \nchi^2)^{3/2} \equiv J_X(p),
  \label{eq:JX}
\end{equation}
which is exactly the translation factor inside $B_0$ reduced by $A^{-3}$, but with norm mode $p^{3/2}$ as a geometric factor.
As we previously noted, the $A^3$ is one power per \emph{translation}, and the difference from the $A^6$ of \cref{eq:boson_zeroes} is due to the phase and rotation modes already being accounted for in \cref{eq:Zfull}.
By \cref{eq:Aneutral} they carry no power of $A$.
To make the connection with the previous bosonic zero mode computation, the accounting of six zero modes in \cref{eq:trademeasure} factorizes block-by-block into internal rotations (and phase) times translations-with-the-internal-block-removed,
\begin{equation}
  (2\pi)^{-3}\sqrt{\det\mathrm G}
  = \underbrace{\vphantom{J_X}\pi^{-3/2}A^3}_{\text{2 rotations} + \text{phase}}
    \;\times\;
    \underbrace{\pi^{-3/2}J_X(1)}_{\text{3 translations}}.
  \label{eq:gramsplit}
\end{equation}
It is the second factor that \cref{eq:JX} reproduces.
The numerical factors come from both the bosonic measure (factors of $\sqrt{\pi}$) and Nambu-doubled space (factors of $\sqrt{2}$); matching and carrying these through is crucial to getting an accurate result as we get in \cref{eq:Irho} below. 

In the rest of this section, we will use the Berezinian along with the transverse integral due to its dependence on the norm mode $p$ which highlights that the frame-independent object must be their product (the fiber density).
Since \cref{eq:JX} is pure body, $\mathcal H$ of \cref{eq:Hweight} below, which carries $J_X$ inside it, \emph{is} that fiber density. 
This is precisely why it is the object with the functional form $H_0(q) + d\,\varkappa(q)$.
Crucially, $\Ber^{\mathrm{body}}$ carries no Grassmann content, so the measure contributes no $d$-term of its own.
Neither statement needs the covariant frame.
Nonetheless, one can carry out the calculation of the covariant Berezinian as a check (not shown).
The end result is that it differs by a factor of $q/p$ per translation zero mode, $\Ber^{\mathrm{cov}} = \Ber^{\mathrm{body}}(q/p)^3$.
This is compensated by the integral over transverse fluctuations, leaving the fiber density unchanged.

\subsection{The fiber density and the Ward reduction}
\label{sec:wardreduction}

Integrating the transverse fluctuations and translations at fixed $\Zc$ defines the fiber density,
\begin{equation}
  \mathcal H(\Zc)  = J_X(p)\int \mathcal D\eta\,\mathcal D\xi\;
    e^{-E_\perp[\Zc;\,\eta,\xi]}
  = H_0(q) + d\,\varkappa(q),
  \label{eq:Hweight}
\end{equation}
where $J_X$ is the collective measure factor \cref{eq:JX}. 
The second equality is forced by invariance as we discussed in \cref{sec:supermatrix}.
The translation trade contributes $\pi^{-3/2}J_X(p)\,d^3X$ by \cref{eq:gramsplit} (we keep the factor $\pi^{-3/2}$ outside $\mathcal H$ by convention, so that $\mathcal H$ is the bare fiber density).

This quantity is one step away from the fluctuation determinant we already computed if we let $\zeta_m$ and $\bar\zeta_m = 0$ in the body frame. 
In fact, in the body frame and without fermionic zeros, we can expand around the saddle point $p=1$,
\begin{equation}
  \begin{aligned}
  \label{eq:Drelrestored}
  \int_{\mathcal J_\downarrow}\!\! p\, dp \; H_0(p)\, e^{-E(p)}
 & = \int_{\mathcal J_\downarrow}\!\! dp \; H_0(p)\, e^{-E(p)}\bigl(1 + \mathcal O(w^2)\bigr) \\
 & = \frac{2\sqrt\pi}{A}\, J_X(1)\, \Drel\, e^{-S_I}\bigl(1 + \mathcal O(w^2)\bigr),
\end{aligned}
\end{equation}
up to phases and the $\mathcal J_\downarrow$ descending norm-mode contour discussed in \cref{app:thimble}.
While in \cref{sec:fluctdet}, we do every Gaussian direction together to compute $\Drel$, here we hold the norm direction $p$ back formally and perform it last due to its crucial role in fixing the Ward identities.

In the end, we obtain the integral in \cref{eq:Drelrestored}, allowing us to use our previous results for $J_X(1)$ and $\Drel$.
All that has to be supplied is the measure on the massive mode left over, $p$. 
The field is not unit norm and $p$ characterizes its deviation from its fixed (stiff) value of $\sqrt2A$. 
The scaled value $\sqrt2A$ accompanies each real collective coordinate including, in particular, $p$ in the Jacobian [\cref{eq:trademeasure}]. 
The negative power of $A$ in the constant $2\sqrt{\pi}/A$ derives from this Jacobian (this is crucial for power-counting).
Finally, the explicit $p$ in the measure is $1 + \mathcal O(w)$ across the Gaussian (first line in \cref{eq:Drelrestored}).\footnote{The final result is $\mathcal O(w^2)$ and not $\mathcal O(w)$ due to \cref{eq:half} in \cref{app:thimble} with $\mathcal J_\downarrow$ as the full line. Namely, $\braket{p}-1 \sim \braket{\delta p} \sim \braket{\delta p^2} H_0'(1)/H_0(1)$ with $\braket{ \delta p^2} = - 1/(2S_I)$.}
Therefore, within the saddle-point approximation $p\, dp (\cdots) \mapsto dp (\cdots)$ and we can use the latter in \cref{eq:master} to connect to this result directly.

With that integral established, we can finally begin to saturate our fermionic zero modes.
Restoring $q$ and $d$ to the full expression of $\mathcal H(\Zc)$, we can identify precisely where the zero modes must live; in the body frame they are $\nu$ in $q = p + \nu$ and the two $c$-numbers are Grassmann quadratic forms $N_\pm$ which appear in $d = p N_- + 2N_+ N_-$.
This frame is convenient to expand $q$ and $d$ into $p$ and $N_\pm$ using Grassmann algebra ($\nu^2 = 2N_+ N_-$ and $N_\pm^2 = 0$) to compute
\begin{equation}
\label{eq:bodyexpand}
\begin{split}
  \mathcal H(\Zc) = H_0(p) &+ \big[N_+ + N_-\big]\, H_0'(p) + p\,\varkappa(p)\, N_- \\
  &+ \big[H_0''(p) + p \varkappa'(p) + 2\varkappa(p)\big]\, N_+ N_-.
\end{split}
\end{equation}
As we have shown in the body frame, $J_X$ carries no Grassmann content so we can write out $\mathcal H(\Zc) = J_X(p) D_\perp(\Zc)$ with $D_\perp(\Zc)$ the transverse integral (at the saddle, the transverse fluctuation determinant, with the collective coordinate $\Zc$ held fixed rather than integrated). In this case, we can find the response function for each pair of Grassmann zero modes
\begin{equation}
\Lambda_m(p) = \partial_{\bar\zeta_m}\partial_{\zeta_m} \ln\mathcal H\big|_{\zeta=0}= \partial_{\bar\zeta_m}\partial_{\zeta_m} \ln D_\perp\big|_{\zeta=0}.
\label{eq:Lambdadef}
\end{equation}
These are the coefficients of $N_m$ in $\ln \mathcal H$, so \cref{eq:bodyexpand} gives
\begin{equation}
  \Lambda_+(p) = \frac{H_0'(p)}{H_0(p)}, \qquad
  \Lambda_-(p) - \Lambda_+(p) = \frac{p\,\varkappa(p)}{H_0(p)}.
  \label{eq:lambda}
\end{equation}
We can already see that the Ward identity fixes how the fermionic part must behave: $\Lambda_+(p)$ is exactly a $p$ derivative of $\ln H_0(p)$ regardless of how the transverse fluctuations dress it.
The only freedom allowed by the fermionic response is the channel \emph{asymmetry} $\Lambda_- -\Lambda_+$ (i.e., the function $\varkappa(p)$).
We will find that a channel symmetry in this problem implies $\varkappa(p)\equiv 0$ identically, see \cref{sec:kappa}, leaving us only with $H_0$.

Now insert the probe to compute the density of states.
On the zero-mode sector the density-of-states insertion is exactly $A^2(p - \nu)$. 
The transverse fluctuations will dress both bosonic $p A^2$ and fermionic $-\nu A^2$ pieces, and their choreography under supersymmetry will be crucial for computing the lowest-order one-loop term we identified earlier in \cref{sec:powercounting}.
Again, using $F(t) = H_0(t)e^{-E(t)}$ and $\mathcal W_\varkappa(t) = \varkappa(t)e^{-E(t)}$ as in \cref{sec:supermatrix}, the Berezin calculus \cref{eq:berezin} tells us \emph{exactly} how saturation occurs 
\begin{equation}
\begin{split}
  p\!\int d^4\zeta\,&(p-\nu)[H_0 + d\varkappa]e^{-E} \\
  &= \big[\,4H_0(p) - 2p^2\varkappa(p)\,\big]e^{-E(p)} \\
  &\quad+ \frac{d}{dp}\Big[p^2F' - 4pF + p^3\mathcal W_\varkappa \Big].
\end{split}
  \label{eq:saturation}
\end{equation}
Notice that similar to \cref{sec:supermatrix}, the total derivative term, once integrated over $\mathcal J$ (the entire integration cycle) must vanish due to boundary terms similar to \cref{eq:noinsertion}; however, not even $p=0$ survives this insertion due to an overall power of $p$, even multiplying $F$.
Relating this back to the $\mathfrak m^2 + Q_4$ in \cref{sec:powercounting}, the $-\nu$ piece provides one fermion pair (the $F'$ term) while the unaccounted-for pair is saturated by the structure of $E$ in the covariant form.
Note how much of this structure is imposed by supersymmetry and the nature of the zero-mode manifold; the detailed contents of interactions, while important for $H_0$ and $\varkappa$, are largely irrelevant up until now.
Every derivative of $H_0$ is buried in a total derivative term and annihilated by the fundamental theorem of calculus, and all due to direct Grassmann integration.

The density of states is therefore determined by one remaining integral
\begin{equation}
  I_\rho = \frac12 \int_{\mathcal J_\downarrow} dp\;
  e^{-E(p)}\, A^2 \big[\, 4 H_0(p) - 2 p^2 \varkappa(p) \,\big].
  \label{eq:master}
\end{equation}
The factor $\tfrac12$ is the retarded half-thimble weight (\cref{app:thimble}), and as we discussed, the corrections are $\mathcal O(w^2)$.

We then use \cref{eq:lambda} to factorize the integrand 
$4H_0(p) - 2p^2\varkappa(p) = H_0(p)\big[4 - 2p\,(\Lambda_- - \Lambda_+)(p)\big]$, ensuring all saddle content is in $H_0\,e^{-E}$ of \cref{eq:Drelrestored}. 
Evaluating $\Lambda_\pm$ at the saddle $p = 1$ defines the factor
\begin{equation}
  \varkappa_1 \equiv \frac{\varkappa(1)}{H_0(1)} = \Lambda_-(1) - \Lambda_+(1). 
  \label{eq:kappadef}
\end{equation}
And thus, we have all necessary information to evaluate the integral in the saddle-point approximation using \cref{eq:Drelrestored} for the $p$ integral and \cref{eq:JX} for $J_X(1)$,
\begin{equation}
  \begin{aligned}
  I_\rho
  &= \frac12\cdot 4A^2\Big(1 -\frac{\varkappa_1}{2}\Big)
     \int_{\mathcal J_\downarrow}\!\! dp\; H_0(p)\, e^{-E(p)} \\
  &= 4\sqrt\pi\,A^2\Big(1 -\frac{\varkappa_1}{2}\Big)\,
     \frac{J_X(1)}{A}\,\Drel\, e^{-S_I} \\
  &= 4\sqrt\pi\,A^4\Big(1 -\frac{\varkappa_1}{2}\Big)\,
     n_t^{3/2}(1-\nchi^2)^{3/2}\,\Drel\, e^{-S_I} ,
  \end{aligned}
  \label{eq:Irho}
\end{equation}
with $A^4 = \gamma^2/w^4$ by \cref{eq:amplitude}, the prefactor of $w^{-4}$ in \cref{eq:result} is established.
Every factor that contributed to this is listed in \cref{tab:ledger}.
The density of states then follows as $\rho(0) = \pi^{-5/2} I_\rho$ in normalized units.
Accounting for where each factor of $A$ came from, we obtain
\begin{equation}
  \underbrace{A^3}_{\text{\cref{eq:JX}}}\;\times\;
  \underbrace{A^2}_{\text{probe}}\;\times\;
  \underbrace{A^{-1}}_{\text{\cref{eq:Drelrestored}}}
  \;=\; A^4 \;=\; \frac{\gamma^2}{w^4}.
  \label{eq:Acount}
\end{equation}
The Ward identity fixes the power $w^{-4}$ exactly, just as we argued in \cref{sec:powercounting}.

\subsection{No channel asymmetry, $\varkappa_1=0$}
\label{sec:kappa}

The last quantity to compute is the channel asymmetry of \cref{eq:lambda}.
At the saddle, $\varkappa_1 = \Lambda_-(1) - \Lambda_+(1)$, where the two channels correspond to the two hedgehogs, $m=+1/2$ and $m=-1/2$ (in the body frame we choose the $m=+1/2$ hedgehog for the saddle-point expansion).
This is just the difference of two response functions of the transverse integral [\cref{eq:Lambdadef}].
In order to obtain $\varkappa_1 = 0$, we need to consider both the measure and the spectrum of the transverse fluctuations.
We can already dispense with the contribution from the \emph{measure}, as noted in \cref{sec:frames}, since no $d$-term arises from the Jacobian.
The \emph{spectrum} will also contribute nothing, but that is due to the discrete symmetry of time reversal and charge conjugation.

This channel symmetry indicated by $\varkappa_1 = 0$ is not a priori obvious because we can construct invariants such as $d = pN_- + 2N_+N_-$ which is inherently asymmetric in $N_+$ and $N_-$.
By \cref{eq:bodyexpand}, the coefficients of $N_+$ and $N_-$ in $\mathcal H$ are $H_0'(p)$ and $H_0'(p) + p\varkappa(p)$ respectively.
Because the difference in these coefficients is exactly the form factor $p \varkappa(p)$, any operation on the transverse integral $D_\perp(\Zc)$ that will exchange the roles of $N_+$ and $N_-$ while holding the bosonic body frame $(\sqrt p, 0)$ fixed will imply that $\varkappa(p)\equiv 0$.
Operationally, the symmetry is just complex conjugation on $D_\perp$ and it relies on a spin-independent and scalar disorder potential.

The full proof of this is in \cref{app:kappa}. We sketch out the proof here for intuition.
First, we note that every parameter in the integral is explicitly real except for the Weyl operator and the Kramers doublet $\ph_{\pm1/2}$ of \cref{eq:doublet}.
In the bosonic sector complex conjugation of $D_\perp$ becomes a trivial operation; in the Nambu doubled space, it corresponds to charge conjugation, where $K \to \bar K$ and we exchange Nambu sectors with $\tau_1$.
We call this entire antiunitary operation $\mathsf C$.
This therefore leaves the entire bosonic sector invariant, including the body frame $(\sqrt p, 0)$.
On the other hand, the fermion sector has not been Nambu doubled, and we can resolve its Weyl operator via the time-reversal operator that squares to $-1$, namely $i\sigma_y$ times complex conjugation.
We call this antiunitary operation $\mathsf T$.
This operation, acting on the zero-mode background, is exactly the swap $\zeta_+ \leftrightarrow \zeta_-$, hence $N_+ \leftrightarrow N_-$.
Taking the operation on bosons and fermions together gives us the total antiunitary $\Theta = \mathsf C \otimes \mathsf T$.
The coupling between bosons and fermions occurs only through the density--density interaction, which is invariant under $\Theta$ since $\Theta$ preserves the superdensity pointwise.
Since $\Theta$ swaps the channels in the fermion sector while leaving the bosonic sector alone, we can now relate the two responses: $\Lambda_+(p) = \Lambda_-(p)^*$.
All that is left is to show that these two responses are entirely real, which we can see simply with the operator $R_y(\pi)\,\mathsf T$ applied to both the bosonic and fermionic sectors (where $R_y(\pi) = e^{-i\pi J_y}$ is the rotation by $\pi$ about the $y$ axis), and therefore
\begin{equation}
  \Lambda_+(p) = \Lambda_-(p) \quad\text{at every real } p.
  \label{eq:matching}
\end{equation}
This is the channel-matching theorem, and it immediately implies that $\varkappa(p) \equiv 0$ and hence $\varkappa_1 = 0$.
This is crucially related to time-reversal symmetry in the original problem.
If we wanted to break it, an easy way would be to introduce magnetic disorder; generically $\varkappa_1 \neq 0$.

We can use numerics to confirm this identity across the full grid of domain radii and angular cutoffs ($R \in \{40,50,60\}\,\xi$, $j_{\max} \in \{4,6,8,10\}$).
Numerically, this allows us to extrapolate to infinite cutoffs and obtain the common channel value $\Lambda_\pm(1) = -2.2835(49)$.
And indeed, the responses agree up to numerical accuracy, $|\varkappa_1| \le 2\times10^{-14}$.
The agreement matches what the antiunitary operator $\Theta$ predicts: the fermionic channels $+$ and $-$ are equal when we sum over all sectors of the magnetic quantum number $M$ [\cref{eq:Moperator}]. Further, across the mixed boson--fermion blocks, the fermionic $+$ channel in bosonic sector $M$ equals the fermionic $-$ channel in bosonic sector $-M$, matching their response functions once $\pm M$ sectors are paired (the conjugate-block pairing of \cref{app:kappa}). The self-conjugate $M=0$ block has a channel difference that vanishes on its own: $+$ and $-$ canceling to $4\times 10^{-16}$.
To test whether we can resolve a finite $\varkappa_1$ when we break $\Theta$, we compute $\varkappa_1 = +0.27$ once a time-reversal-symmetry-breaking term is inserted.

With $\varkappa_1 = 0$ in \cref{eq:kappadef}, we do not have to consider this term any longer.
Without other quantities to compute, we can assemble the pieces to obtain the density of states.

\subsection{The assembled prefactor}
\label{sec:assembly}

All that is left is arithmetic. We have established $\varkappa_1 = 0$ by \cref{sec:kappa} and performed the integral over the norm mode $p$ in \cref{eq:Drelrestored}. 
Performing the $p$ integral restored the fluctuation determinant from \cref{sec:fluctdet} (measured in \cref{fig:fluct} and shown in reduced Fredholm determinant form in \cref{eq:fredholm}).
\Cref{tab:ledger} shows all of the contributions to the density of states as we have computed them, first assembling \cref{eq:Irho}, and then, from that, the overall coefficient $\mathcal A$.

\begin{table}[t]
  \caption{\textbf{Contributions to the density of states}. Rows in the first block multiply to $I_\rho = \pi^{5/2}\rho(0)$, \cref{eq:Irho}; the remaining rows establish \cref{eq:assembly}, using $A^4 = \gamma^2w^{-4}$ [\cref{eq:amplitude}] and $\varkappa_1 = 0$ [\cref{sec:kappa}]. Powers of $A$ multiply to $A^4$.}
\label{tab:ledger}
\begin{ruledtabular}
\begin{tabular}{lll}
factor & value & fixed in \\
\colrule
translation Jacobian & $J_X(1) \sim A^3$ & \cref{eq:JX} \\
doublet measure & $p\,dp\,d^4\zeta \sim A^0$ & \cref{eq:Zfull,eq:Aneutral} \\
probe & $A^2(p-\nu) \sim A^2$ & \cref{eq:master} \\
Berezin saturation & $4\,(1-\varkappa_1/2)$ & \cref{eq:saturation,eq:kappadef} \\
$p \to \vartheta$ Jacobian & $2\sqrt\pi/A \sim A^{-1}$ & \cref{eq:Drelrestored} \\
fluctuation determinant & $\Drel$ & \cref{eq:Drel,eq:Drelnum} \\
half-thimble & $\tfrac12$ & \cref{app:thimble} \\
\colrule
\multicolumn{3}{c}{$I_\rho
 = 4\sqrt\pi\,A^4\big(1-\tfrac{\varkappa_1}{2}\big)\,
   n_t^{3/2}(1-\nchi^2)^{3/2}\,\Drel\,e^{-S_I}$}\\
\colrule
trade normalization & $\pi^{-3/2}$ & \cref{eq:gramsplit} \\
translation volume & $L^3$ & $\int d^3X$ \\
density of states trace & $1/2\pi L^3$ & \cref{eq:dos,eq:insertion} \\
saddle multiplicity & $2$ & \cref{sec:saddle} \\
phases & net $+1$ & \cref{app:thimble} \\
\colrule
\multicolumn{3}{c}{$\mathcal A
 = \dfrac{4}{\pi^2}\,\gamma^2\,n_t^{3/2}(1-\nchi^2)^{3/2}\,\Drel
 = 27.47$}\\
\end{tabular}
\end{ruledtabular}
\end{table}

Fully assembled, the density of states is therefore
\begin{equation}
\begin{aligned}
  \rho(0)
  &= \frac{\mathcal A}{\hbar v\, \xi^2}
     \Big(\frac{\hbar v}{W\xi}\Big)^{4}
     \exp\!\Big[-s^*\Big(\frac{\hbar v}{W\xi}\Big)^{2}\Big]
     \big(1 + \mathcal O(w^2)\big), \\
  \mathcal A
  &= 2\cdot\frac{2}{\pi^2}\, \gamma^2\, n_t^{3/2}
     (1-\nchi^2)^{3/2}\, \Drel\,\Big(1 - \frac{\varkappa_1}{2}\Big) \\
  &= 27.47 \pm 0.08,
\end{aligned}
\label{eq:assembly}
\end{equation}
which is \cref{eq:result}.
The leading factor of $2$ is the saddle multiplicity: the well and barrier instantons of \cref{sec:saddle} are exchanged by parity, so their half-thimble contributions to the retarded sum are equal and add.
The retarded factor $\tfrac12$ inside is the spectral weight of the zero crossing --- the density of states samples the boundary value $1/(x+i0^+)$, which carries exactly half the Stokes discontinuity of the norm-mode thimble --- and the phases are all tallied in \cref{app:thimble}.
The error is dominated by $\Drel$; every other input is an analytic identity or a quadrature-certified integral, and the assembled $\mathcal A$ shifts by only $1.2\times10^{-4}$ (absolute, with $\Drel$ held at its extrapolated value) between domain radii $R = 40$ and $60$.

Two remarks bound the scope of \cref{eq:assembly}.
First, while the analytics require only smoothness and the $r^{-4}$ tail, the values $(s^*, \mathcal A)$ are kernel-specific.
Second, two reasonable explanations can be given for the $0.76$ factor difference between numerical and analytic data in \cref{fig:dos}. One is analytic: the $\mathcal O(w^2)$ correction could account for it. In fact, the nearly flat factor $0.76$ is equivalent to a correction coefficient $c \approx -(0.4$--$0.6)$ in $\rho \to \rho\,(1 + c\,w^2)$ at the accessible $w^2 \approx 0.4$--$0.6$. The other explanation is numeric: the $0.76$ factor could also be due to systematics within the numerical data approaching a continuum limit; indeed, increasing $\xi$ relative to the lattice spacing used in the discretization moves $\rho(0)$ numerically up and closer to this value \cite[see][Fig.~3(c)]{WilsonPixley2020a}. The data presented here was the \emph{closest} to the $\xi\gg a$ regime ($\xi=2a$) from that numerical campaign.

\section{Conclusion and outlook}
\label{sec:conclusion}

The zero-energy density of states of a disordered three-dimensional Weyl semimetal is nonzero at every finite disorder strength, \cref{eq:result}, with no fluctuation catastrophe available to restore $\rho(0) = 0$: the requisite fermionic zero modes do not exist [\cref{eq:kerK}, with $57\%$ margin], and the fluctuation determinant is pinned at $\mathcal O(1)$ by the supersymmetric anchor \cref{eq:anchor}, verified to $2.5$ parts in $10^6$ of the raw spectral sums.
An exact Ward identity lets us saturate fermionic zero modes and leads to the one-loop result and $w^{-4}$ prefactor.
The one term that could have corrected this is zero by the channel-matching theorem [\cref{eq:matching}].
The perturbative semimetal--metal critical point, as measured by the density of states, is therefore avoided \cite{PixleyDasSarma2016a, WilsonPixley2020a, PiresLopes2021, PixleyWilson2021}, in the same sense in which Lifshitz tails round band edges \cite{Yaida2016}.

However, there is one major caveat to our story.
We have established that the \emph{mean} density of states is finite and smooth through the perturbative transition.
This does not rule out that typical observables could show some form of criticality.
Such quantities, including the typical density of states and conductivity at the node, could in principle retain some critical signatures at a finite disorder strength indicative of a critical point even though $\rho(0)$ does not.
Typical observables are usually blind to rare configurations, the quintessential example being Lifshitz tails \cite{Yaida2016}.
However, the power-law rare states in this paper are categorically different, since their amplitude dies off like $1/r^2$ instead of exponentially.
They could hybridize with each other and therefore affect typical quantities in ways that exponentially localized rare states cannot.
In fact, current numerical evidence suggests that these states do exactly that.
The nodal conductivity tracks the rare-region-dominated $\rho(0)$ rather than any critical form, with $\sigma(0)\propto\rho(0)^2$ observed numerically \cite{FuPixley2024} (and in tension with the constant diffusivity expected from a dilute gas of scatterers).
Further, a lattice study of dilute vacancies (resonant point defects rather than the smooth potential considered here) shows the same pattern: the dc conductivity tracks the defect-induced nodal peak in $\rho(E)$, dipping at the hybridized resonances where the enhanced density of states is offset by a strongly suppressed diffusivity \cite{SantosPiresVianaParenteLopes2022}.
On the other hand, power-law states occur \emph{at} the mobility edge of a band separating Lifshitz tails from diffusive states; there, multifractal states appear and transport acquires a characteristic form.
The nodal point of a Weyl semimetal might realize exactly that kind of physics up until a critical point.
As a result, settling the typical sector analytically remains open.

Lastly, future work can address the $\mathcal O(w^2)$ coefficient, a dilute (Coulomb) gas of instantons in the spirit of Ref.~\cite{Gurarie2017}, and other types of disorder.

\begin{acknowledgments}
I thank M.~S.~Foster, B.~Roy, P.~W.~Brouwer, and L.~Radzihovsky  for discussions, and J.~H.~Pixley, P.~Chandra, and A.~A.~Allocca for discussions and collaboration on related work.
I owe a particular debt to M.~Buchhold, S.~Diehl, and A.~Altland for discussions of the instanton problem their work opened.
This work was supported by the National Science Foundation under NSF CAREER Grant No.~DMR-2238895.
Part of this work was performed at the Aspen Center for Physics, which is supported by National Science Foundation Grant No.~PHY-2210452.
Portions of this research were conducted with high-performance computational resources provided by Louisiana State University (\url{http://www.hpc.lsu.edu}).
This research was done using services provided by the OSG Consortium \cite{Pordes2007,Sfiligoi2009,OSPool,OSDF}, which is supported by the National Science Foundation awards \#2030508 and \#2323298.
\end{acknowledgments}

\section*{Data Availability}
All numerics reported here --- the instanton profile of \cref{sec:saddle}, the zero-mode ledger, and the Fredholm determinant of \cref{eq:fredholm} --- were carried out in the Julia programming language \cite{BezansonShah2017}.
The code and the data that support the findings of this article are not yet publicly available but will be deposited in a Zenodo repository; in the meantime they are available from the author upon reasonable request.

\appendix
\crefalias{section}{appendix}
\crefalias{subsection}{subappendix}
\crefalias{subsubsection}{subsubappendix}

\section{Norm and translation modes in the fluctuation determinant}
\label{app:normtrans}

The primes in \cref{eq:Drel} remove mismatched sets: $\mathcal F \equiv K \oplus \bar K$ has four real zero directions (phase, two rotations, norm) while $H_B = \mathcal F + \mathfrak A$ has six (the same phase and rotations, plus three translations).
The symmetry directions of phase and rotation as well as the $H_{\mathrm{free}}$ normalization cancel in the ratio; the two remaining unmatched sets, the norm mode and the translations, produce factors in front of the Fredholm determinant in \cref{eq:fredholm}.

On the norm direction the anomalous block is strictly negative, $\braket{\Phi_N, \mathfrak A \Phi_N}/\braket{\Phi_N,\Phi_N} = -2\gamma\braket{\hat n, \bC * \hat n} = -4s^*\!/\gamma$: the fermionic zero lifted into the single negative mode of $H_B$.
In the basis of the normalized norm mode $\Phi_N/\sqrt2$ and its complement, the norm row and column of $H_B$ are eliminated by a Schur complement,
\begin{equation}
\begin{aligned}
  H_B &=
  \begin{pmatrix}
    -\tfrac{4s^*}{\gamma} & \sqrt2\, U(y)\hat\Phi^\dagger(y) \\
    \sqrt2\, U(x)\hat\Phi(x) & \mathcal F + \mathfrak A
  \end{pmatrix} \\
  &\simeq
  \begin{pmatrix}
    -\tfrac{4s^*}{\gamma} & 0 \\
    0 & \mathcal F + \mathfrak A_{\mathrm{eff}}
  \end{pmatrix},
\end{aligned}
\label{eq:normschur}
\end{equation}
where $\simeq$ stands for $H_B = L\,(\cdots)\,L^\dagger$ with the unit-determinant row reduction
\[
  L = \begin{pmatrix}
    1 & 0 \\
    -\tfrac{\gamma}{2\sqrt2\, s^*}\, U(x)\hat\Phi(x) & 1
  \end{pmatrix},
\]
and $\mathfrak A_{\mathrm{eff}} = \mathfrak A + \frac{\gamma}{2s^*}\, U(x)\hat\Phi(x)\, U(y)\hat\Phi^\dagger(y)$ is \cref{eq:Aeff}.
This extracts the factor $4s^*\!/\gamma$ of \cref{eq:fredholm}, and now the lower-right block $\mathcal F + \mathfrak A_{\mathrm{eff}}$ no longer includes the fermionic zero mode.
We can now apply $G_I' = \mathcal F^{-1}$, the inverse of $\mathcal F$ with the zero modes of $K$ removed, so that $\mathcal F + \mathfrak A_{\mathrm{eff}} = \mathcal F\,(1 + G_I'\mathfrak A_{\mathrm{eff}})$.
With this elimination, we suspect it can be \emph{proved} that $G_I' \mathfrak{A}_\mathrm{eff}$ is trace-class (numerically, $\|G_I'\mathfrak A_{\mathrm{eff}}\|_1 = 11.3$), but we leave that to future work.

All that is left is to remove the three translation zero modes.
These are zeros of the bosonic numerator $\mathcal F + \mathfrak A_{\mathrm{eff}}$ but not of the invertible fermionic denominator $\mathcal F$, so removing them leaves a finite kernel metric.
We define the tangents of the translation mode as $t_a = \partial_a\hat\Phi$. 
However, we have already projected out the phase and rotation directions from both the fermionic and bosonic sectors, so we also need to define $t_a^\perp$ by projecting off those directions. 
Note that $H_B t_a^\perp = 0$ exactly.
With both $t_a$ and $t_a^\perp$, the resulting metrics are
\begin{equation}
\begin{aligned}
  \mathsf G_{ab} &= \braket{t_a^\perp, t_b^\perp}
    = 2 n_t (1-\nchi^2)\, \delta_{ab}, \\
  \mathsf H_{ab} &= \braket{t_a, \mathcal F t_b}
    = h_T\, \delta_{ab}.
\end{aligned}
  \label{eq:kernelmetric}
\end{equation}
This leads into \cref{eq:fredholm}, $\Drel^{-2} = \frac{4s^*}{\gamma}\, (\det\mathsf G/\det\mathsf H)\, |{\Det}'(1 + G_I' \mathfrak A_{\mathrm{eff}})|$.
The factor $(1-\nchi^2)$ per direction is the same collective Schur factor as in $B_0$ --- the translation tangents overlap the internal ones, $\braket{\ph, \partial_z\ph} = 2is^*\!/3\gamma$ --- while $\mathsf H = -\braket{t_a, \mathfrak A t_b} = \gamma\braket{\partial_a \hat n, \bC * \partial_b \hat n}$ by \cref{eq:goldstone}, positive because $\hat{\bC}(k) > 0$.
Numerically ($E_\xi = \hbar v/\xi = 1$): $4s^*\!/\gamma = 2.1183$, $2n_t = 1.3446$, $h_T = 0.70580$; the front factors contribute $2.069$ in the log against $\ln|{\Det}'| = -2.328$ at $R = 60\,\xi$, assembling to $-2\ln\Drel = -0.259$ against $-0.2606(60)$ of \cref{fig:fluct} from the route that explicitly subtracts the bosonic and fermionic log determinants [\cref{eq:LBLF}]. 
This shows that these two algebraically distinct methods for computing the fluctuation determinant agree; subtracting the raw determinants is numerically equivalent to a Fredholm determinant of $G_I'\mathfrak A_\mathrm{eff}$.

Using the standard inequalities of trace-class operators, we note
\begin{equation}
  |\Det(1+T)| = \prod_j |1+\mu_j| \le e^{\sum_j |\mu_j|} \le e^{\|T\|_1}
  \label{eq:traceineq}
\end{equation}
(removing the prescribed $\mu = -1$ factors, i.e., the translation zero modes, only strengthens this).
Hence
\begin{equation}
  -2\ln\Drel \le \ln\frac{4s^*}{\gamma}
  + 3\ln\frac{2n_t(1-\nchi^2)}{h_T}
  + \|G_I'\mathfrak A_{\mathrm{eff}}\|_1.
  \label{eq:tracebound}
\end{equation}
Now note that \cref{eq:tracebound} is independent of $S_I$ (by construction since every term originates from the bare $\mathcal O(w^0)$ integral).
No super-exponential suppression can hide in the fluctuation determinant, given the trace-class nature of $G_I'\mathfrak A_\mathrm{eff}$, supported by the numerics.

\section{The channel-matching theorem}
\label{app:kappa}

In this appendix we prove \cref{eq:matching}.
In the body frame [\cref{eq:bodygauge}], as we showed in the main text in \cref{eq:JX}, the Jacobian $J_X(p)$ has no Grassmann content.
In particular, $\mathcal H(\Zc) = J_X(p)\,\Dperp$, and thus all of the Grassmann content lives in $\Dperp$. 
We therefore expand $\ln\Dperp$ using Grassmann algebra as
\begin{equation}
  \ln \Dperp = L_0(p) + \Lambda_+(p)\,N_+ + \Lambda_-(p)\,N_- + L_2(p)\,N_+N_-.
  \label{eq:Dperpexpand}
\end{equation}
In this expression the channel matching reduces to showing the symmetry $\Lambda_+(p) = \Lambda_-(p)$.
This forces $\varkappa \equiv 0$ and with it $\varkappa_1 = 0$.
We prove it by exhibiting an operation on the transverse integral that exchanges the two pairs, $N_+ \leftrightarrow N_-$, while leaving the body $(\sqrt p, 0)$ --- and everything else --- alone.

\emph{Time reversal.}---The time-reversal operator $\mathsf T$ is the standard one for spin $\tfrac12$, with $\mathsf T^2 = -\mathds 1$, and it transforms spinor wave functions antiunitarily as
\begin{equation}
  \mathsf T u = i\sigma_y u^*.
  \label{eq:Tdef}
\end{equation}
The identity $\sigma_y\bm\sigma^*\sigma_y = -\bm\sigma$ makes the Weyl operator $\mathsf T$-real,
\begin{equation}
  (i\sigma_y)^\dagger\,(-i\bm\sigma^*\!\cdot\nabla)\,(i\sigma_y) = i\bm\sigma\cdot\nabla,
  \label{eq:TKweyl}
\end{equation}
and every other parameter or function in the action \cref{eq:E} (e.g., $\gamma$, $A$, $\sqrt p$, $U(x)$, $\bC(x-y)$) is real.
The only complex parameters are the Weyl operator and the doublet itself, and the doublet is closed under $\mathsf T$: fixing the phase convention $\ph_{-1/2} = \mathsf T\,\ph_{+1/2}$ for the partner of \cref{sec:saddle}, $\mathsf T^2 = -\mathds 1$ gives $\mathsf T\,\ph_{-1/2} = -\ph_{+1/2}$, i.e.,
\begin{equation}
  \bar\ph_{+1/2} = -i\sigma_y\,\ph_{-1/2},
  \qquad
  \bar\ph_{-1/2} = +i\sigma_y\,\ph_{+1/2}.
  \label{eq:doubletconj}
\end{equation}

\emph{Conjugation is the whole operation.}---The coefficients in \cref{eq:Dperpexpand} can be computed from the fluctuation integral and all objects it contains (parameters and operators). In addition to the free part, the interactions in \cref{eq:Eexpand} can be computed as sums of Wick contractions using those very same objects.
Conjugating the objects within the transverse integral therefore amounts to conjugating $\Dperp$ itself,
\begin{equation}
  \Dperp\big[\text{objects}^*\big] = \Dperp\big[\text{objects}\big]^*,
  \label{eq:conjmove}
\end{equation}
where the asterisk conjugates the functions $L_0$, $\Lambda_\pm$, $L_2$ and leaves Grassmann generators alone.
The theorem follows from evaluating the left-hand side via an antiunitary operation that is charge conjugation on the boson sector while being time reversal on the fermion sector (in effect swapping the $m=\pm1/2$ channels for fermions while leaving them alone for bosons).

\emph{Bosons: Charge conjugation.}---At the saddle point the bosonic field has $\bar \varphi_I = \varphi_I^\dagger$ and we add fluctuations such that $\delta \bar \varphi = \delta \varphi^\dagger$, staying on the contour defined by the instanton (except for transverse negative modes which we collect in \cref{app:thimble}).
This is constructed in such a way that the Nambu spinor $\delta\Phi = (\delta\varphi, \delta\bar\varphi)^T$ of \cref{eq:HB} is manifestly symmetric under charge conjugation $\mathsf C$ defined by $\mathsf C \delta\Phi = \tau_1 \delta \Phi^* = \delta \Phi$.
All that remains is ensuring the quadratic form \cref{eq:HB} is also symmetric
\begin{equation}
  \mathsf C H_B \mathsf C = \tau_1\, H_B^*\,\tau_1 = H_B,
  \label{eq:HBnambu}
\end{equation}
where we have used the saddle's $\bar\varphi_I = \varphi_I^\dagger$. 
(This applies to both the doubled Weyl operator $K \oplus \bar K$ and the operator $\mathfrak A$.)
Objects such as the boson density $\bar\varphi_x\varphi_x = |\varphi_x|^2$ and the Nambu spinor of the body $A\sqrt p\,(\ph_{+1/2}, \bar\ph_{+1/2})^T$ are invariant under charge conjugation $\mathsf C$.
Likewise, the removed zero modes and the constraints they imply for the transverse fluctuations are invariant under charge conjugation.
So the boson sector passes through \cref{eq:conjmove} untouched, and in particular $p$ does.
The lack of an operation on the spin degree of freedom differentiates $\mathsf C$ from the fermionic operation, $\mathsf T$.

\emph{Fermions: Time-reversal.}---A Grassmann bilinear has no reality of its own, so in the fermion sector the conjugated parameters must instead be rotated back.
The fermions enter \cref{eq:Eexpand} only through $\int_x\bar\chi K\chi$ and the scalar density $n_F = \bar\chi_x\chi_x$, so every operator between $\bar\chi$ and $\chi$ is built from two ingredients alone --- the Weyl operator, and real multiplication ($U$ and the $\bC$-smeared densities).
The substitution
\begin{equation}
  \chi_x = i\sigma_y\,\chi'_x,
  \qquad
  \bar\chi_x = \bar\chi'_x\,(i\sigma_y)^\dagger,
  \label{eq:fermsub}
\end{equation}
does not change the measure,
\begin{equation}
  \mathcal D[\bar\chi,\chi] = \mathcal D[\bar\chi',\chi'],
  \label{eq:fermmeasure}
\end{equation}
since it has unit Berezinian [$\det(i\sigma_y) = 1$].
\Cref{eq:TKweyl} along with the reality of $U$ implies $\bar\chi\,K^*\chi = \bar\chi'\,K\,\chi'$, and thus we can absorb the conjugation of $K$ into the redefinition of $\chi$.
Time reversal also maps the doublet space onto itself [\cref{eq:doubletconj}] while preserving orthogonality, hence $\mathsf T$ maps the transverse domain onto itself.
The swap comes about via exactly this action, since we have fixed the Grassmann numbers associated with the Kramers doublet; by \cref{eq:doubletconj},
\begin{equation}
\begin{split}
  A\big(\zeta_+\bar\ph_{+1/2} + \zeta_-\bar\ph_{-1/2}\big)
  &= i\sigma_y\, A\big(\zeta'_+\ph_{+1/2} + \zeta'_-\ph_{-1/2}\big),\\
  (\zeta'_+,\, \zeta'_-) &= (\zeta_-,\, -\zeta_+).
\end{split}
\label{eq:swap}
\end{equation}
Similarly, $(\bar\zeta'_+, \bar\zeta'_-) = (\bar\zeta_-, -\bar\zeta_+)$. 
This is the key step, where $i\sigma_y$ has swapped $\zeta_-$ and $\zeta_+$ for both the barred and unbarred Grassmann variables, rotating the background by the same $i\sigma_y$ as the fluctuation, so the single substitution \cref{eq:fermsub} handles both.
This directly implies that the occupations swap as well,
\begin{equation}
  N_+ \longmapsto N_-,
  \qquad
  N_- \longmapsto N_+.
  \label{eq:Nswap}
\end{equation}
Consequently, $\nu$ and $q = p + \nu$ do not change, while $d$ does change, $d = pN_- + 2N_+N_- \mapsto pN_+ + 2N_+N_-$.
(It is worth also highlighting that \cref{eq:fermmeasure} implies that the zero modes maintain their measure $d^4\zeta' = d^4\zeta$.)

We can treat the two sectors differently because they interact through the local superdensity: $n_x = \bar\varphi_x\varphi_x + \bar\chi_x\chi_x$ is the sole coupling in \cref{eq:E}, and each term is preserved pointwise by this operation: $|\varphi_x|^2$ trivially and $\bar\chi'_x(i\sigma_y)^\dagger(i\sigma_y)\chi'_x = \bar\chi'_x\chi'_x$ by unitarity.
Transforming both the bosonic and fermionic sectors in \cref{eq:conjmove}, we obtain
\begin{equation}
  \Dperp^*(p;\, N_+, N_-) = \Dperp(p;\, N_-, N_+),
  \label{eq:thetamove}
\end{equation}
and hence it follows that
\begin{equation}
  \Lambda_+(p) = \Lambda_-^*(p).
  \label{eq:Lambdaconj}
\end{equation}
These two operations are exactly the antiunitary $\Theta = \mathsf C \otimes \mathsf T$ quoted in \cref{sec:kappa}.
If we applied only time reversal $\mathsf T$ on both sectors, the boson body would also rotate from $(\sqrt p, 0)$ to $(0, \sqrt p)$, leaving \cref{eq:Dperpexpand} invariant; explicitly, $\mathsf T$ will swap $\Lambda_+(p) N_+ \leftrightarrow \Lambda_-(p) N_-$ but $\ln D_\perp$ remains the same after this operation and hence we obtain no information relating $\Lambda_+$ and $\Lambda_-$. 
The solution is to use $\mathsf C$ on the bosons so that only $N_+\leftrightarrow N_-$.

\emph{Reality closes the proof.}---All that remains is to show that $\Lambda_\pm(p)$ are real.
Compose $\mathsf T$, now applied identically to bosons and fermions, with the rotation $R_y(\pi)$. 
This is a symmetry, since $U$ and $\bC$ are isotropic. 
As a substitution the unitary parts of these operators combine into the purely orbital rotation $R_y(\pi)\,i\sigma_y = e^{-i\pi L_y}$, manifestly measure-preserving.
On the doublet this new operator is channel-\emph{diagonal} with $R_y(\pi)\mathsf T\,\ph_m = \ph_m$, so this operation restores every conjugated parameter with every $N_\pm$ fixed.
\Cref{eq:conjmove} then reads $\Dperp^*(p; N_+, N_-) = \Dperp(p; N_+, N_-)$, from which we deduce that every coefficient is real at real $p$, and combined with \cref{eq:Lambdaconj},
\begin{equation}
  \Lambda_+(p) = \Lambda_-(p)
  \quad \text{at every real } p > 0.
  \label{eq:matchingapp}
\end{equation}
By \cref{eq:lambda}, $p\,\varkappa(p) = H_0(p)\big[\Lambda_-(p) - \Lambda_+(p)\big] = 0$.
Lastly, since $\Dperp$ is analytic in $p$, $\varkappa \equiv 0$ more generally --- in particular along the contour $\mathcal J$ --- and at the saddle $\varkappa_1 = \Lambda_-(1) - \Lambda_+(1) = 0$.

The retarded shift $i0^+$ in \cref{eq:action} is the one number, albeit infinitesimal, that complex conjugation conjugates and nothing conjugates back.
The theorem is proven for $\omega = 0$ without any shift, which is inconsequential here because the transverse integral $\Dperp$ does not contain any zero modes.
In the $\hat M$-locked sectors of \cref{sec:fluctdet}, $\Theta$ maps each sector onto its conjugate partner rather than onto itself, so \cref{eq:matchingapp} holds between conjugate blocks, not within blocks with the same magnetic quantum number $M$. 
This is resolved in the numerics quoted in \cref{sec:kappa}, whose per-sector channel responses are genuinely unequal and equalize only when conjugate sectors are summed.
This theorem depends solely on time reversal; scalar disorder is inherently time-reversal invariant, whereas magnetic or other $\mathsf T$-odd couplings will spoil the theorem and generically create a nonzero $\varkappa_1$.

\section{The thimble and the phase ledger}
\label{app:thimble}

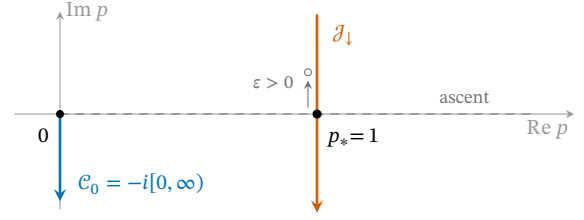
\begin{figure}
  \centering
  \begin{tikzpicture}[>=stealth, line cap=round]
    \draw[black!55, dashed, line width=0.5pt] (0,0) -- (6.3,0);
    \node[black!55, anchor=south] at (5.35,0.03) {\scriptsize ascent};
    \draw[->, black!40, thin] (-0.6,0) -- (6.8,0)
      node[below left=-1pt] {\footnotesize $\mathrm{Re}\,p$};
    \draw[->, black!40, thin] (0,-1.25) -- (0,1.35)
      node[right=-1pt] {\footnotesize $\mathrm{Im}\,p$};
    \draw[okblue, line width=1.0pt, ->] (0,0) -- (0,-1.15);
    \node[okblue, anchor=west] at (0.12,-0.92)
      {\footnotesize $\mathcal C_0 = -i[0,\infty)$};
    \draw[okverm, line width=1.0pt, ->] (3.4,1.3) -- (3.4,-1.3);
    \node[okverm, anchor=west] at (3.5,1.05)
      {\footnotesize $\mathcal J_{\downarrow}$};
    \draw[->, black!55, thin] (3.28,0.09) -- (3.28,0.44);
    \draw[black!55, thin, fill=white] (3.28,0.55) circle (1.3pt);
    \node[black!55, anchor=east] at (3.2,0.42) {\scriptsize $\varepsilon > 0$};
    \filldraw (3.4,0) circle (1.5pt);
    \node[anchor=north west] at (3.42,-0.07) {\footnotesize $p_*{=}\,1$};
    \filldraw (0,0) circle (1.3pt);
    \node[anchor=north east] at (-0.04,-0.07) {\footnotesize $0$};
  \end{tikzpicture}
  \caption{\textbf{Norm-mode integration contours at $\omega=0$.}
The rotated original contour $\mathcal C_0$ ends at $p = 0$; the thimble $\mathcal J_\downarrow$ ($p = p_* - iy$) is the vertical steepest-descent line through the saddle, traversed downward.
The saddle's steepest-ascent path (dashed) is the real axis and passes through the endpoint of $\mathcal C_0$. 
In this configuration the retarded boundary value picks up half the saddle's discontinuity, \cref{eq:half}.
For $\varepsilon > 0$ the saddle lifts off the axis (grey arrow) and $\mathcal C_0$ misses the thimble entirely.}
  \label{fig:thimble}
\end{figure}

The factor $\tfrac12$ entering \cref{eq:master} and \cref{eq:result} is determined solely by the norm of the instanton, which we obtain by the contour deformation in \cref{sec:supermatrix} [\cref{eq:Zfull}].
Let $z \in \mathbb C^2$ be the amplitude of the instanton doublet and $p = z^\dagger z$ as in \cref{sec:supermatrix}; integrating the unit $S^3$ reduces the measure to $p\,dp$ (the insertion adds one more power of $p$).
On the rotated original cycle $\mathcal C_0 = -i[0,\infty)$ the reduced action is $E_b(p) = S(2p - p^2) + bp$ with $S = S_I$ and $b = (\omega + i\varepsilon)A^2$.
The saddle sits at $p_* = 1 + b/2S$, and the Lefschetz thimble through it \cite{Witten2011} (the steepest-descent contour, on which $e^{-E_b}$ decays as a pure Gaussian) is the vertical line
\[
  \mathcal J_\downarrow:\quad p = p_* - iy, \qquad y \in \mathbb R,
\]
traversed downward [\cref{fig:thimble}].
Both contour integrals are error functions or Gaussians
\begin{equation}
  \begin{aligned}
    J_k & = \int_{\mathcal J_\downarrow} p^k e^{-E_b} = -i \sqrt{\frac{\pi}{S}} e^{-S p_*^2} h_k(p_*),\\
    I_k & = \int_{\mathcal C_0} p^k e^{-E_b} = \frac12 \erfc(-i\sqrt{S} p_*) J_k + R_k,
  \end{aligned}
  \label{eq:JkIk}
\end{equation}
where $R_k$ come from the $p=0$ endpoint and $h_k$ represent Gaussian moments (both are real at $b=0$).
By utilizing $\erfc(x) = 2 - \erfc(-x)$, one can show that the thimble leads to a binary weight for $I_k$. 
It is zero on the retarded side when the saddle goes up into the upper half-plane such that $\mathcal C_0$ cannot be deformed to reach it. 
Contrariwise, its weight is one on the other side; this switch is captured smoothly by the universal error function \cite{Berry1989}.
The density of states is computed at exactly the point where the thimble weight switches from zero to one.
Indeed, at $b=0$, the saddle's steepest-ascent curve is up the real axis from $p=1$ and including the endpoint of $\mathcal C_0$ [$p=0$, \cref{fig:thimble}]. 
This is where the same error function gives (utilizing \cref{eq:JkIk} and $\erfc(-i \tilde x)=1+i\erfi(\tilde x)$ at $b=0$)
\begin{equation}
  \frac{\Imr I_k}{\Imr J_k} = \frac12 \qquad (k = 0, 1, 2, \ldots).
  \label{eq:half}
\end{equation}
The same physical measure ($k=1$) and density-of-states insertion ($k=2$) both only realize half of the integral over the full thimble $\mathcal J_\downarrow$ due to the saddle's discontinuity.

This simplified model applies to the full problem.
The saddle is a continuous orbit, and the generalized thimble fibers the single descending norm direction over the compact $S^3$ (i.e., the radius is all that enters).
The basic content of \cref{eq:half} originates from the identity $1/(x + i0^+) = \mathrm{PV}\,\tfrac1x - i\pi\,\delta(x)$.
The relevant contribution for the Green's function is $\int dx\, F(x)/(x + i0^+)$ where $x$ is not on the contour, but an eigenvalue that crosses zero.
Therefore, when $F(0)\neq 0$ as we tune $x$ through the zero crossing, we have that $-\Imr$ picks up half of the full discontinuity ($\pi$ instead of the poles' $2\pi$).
The crossing is simple, and we can understand it by parametrizing $x = \epsilon(\lambda)$ where $\epsilon(\lambda)$ can be approximated via first-order perturbation theory in $K_\lambda = i\bm\sigma\cdot\nabla + \lambda U$ giving $\epsilon'(1) = -\gamma\braket{\hat n, \bC * \hat n} = -2s^*/\gamma \ne 0$ (and hence $x \approx (\lambda - 1) \epsilon'(1)$). 
To step back from the parameter $\lambda$, we can explicitly look at the full transverse operator as a function of $p$.
When we do a finite-basis scan on this $p$-dependent operator, no other boson or fermion zero is found on $0 < p < 1$ (the sole bosonic root, $p = 1/3$, is the norm caustic of $E_0$ itself with norm-mode overlap $> 0.9999999$).
The ratio $\tfrac12$ is robust here; smooth transverse factors multiply the boundary values and discontinuity equally, thereby having no effect on it.

As a final bit of bookkeeping, we can account for all phases in the problem. 
\begin{enumerate}
  \item When computing the transverse fluctuations (and excluding the norm mode), we encounter terms like $\det(K\oplus \bar K)^{1/2}/\det(H_B)^{1/2}$ which have phases when the operators (e.g., $H_B$) have negative eigenvalues.
    When the particle--hole-paired symmetry sectors for bosons and fermions are combined, the negative eigenvalues contribute the sign $(\pm i)^{n_-(H_B) - n_-(K\oplus\bar K)}$ (where $n_-(A)$ counts the negative eigenvalues of $A$). 
    This difference in the exponent is measured and vanishes sector by sector.
  \item The full Berezinian has no phase due to the $e^{i\pi/4}$ on the supervector; this is simply due to the phase affecting an equal number of bosonic and fermionic directions. However, the rotation does affect the insertion $\bar \psi \mathsf k \psi \mapsto i \bar \psi \mathsf k \psi$ leading to an extra $+i$ factor.
  \item The norm-mode thimble contributes $e^{-i\pi/2}=-i$.
\end{enumerate}
The above product is unity, so the $-i/2$ from \cref{eq:insertion} gives us $G^+|_\mathrm{saddle} = -\tfrac i 2 e^{-S_I} \times (\mathrm{positive})$, implying that the density of states is real and positive, \cref{eq:dos}.

The ledger above is per saddle.
We have two equally weighted instantons, the well instanton and the barrier instanton, connected by the parity operation.
Since this parity operation commutes with the contour rotation and deformation, these two contributions to the density of states are equal and add, contributing a factor of $2$ to our final expression in \cref{eq:assembly}.

\bibliography{references}

\end{document}